\documentclass[aps,prd,reprint,superscriptaddress,nofootinbib,floatfix,amsmath,amssymb]{revtex4-2}
\pdfoutput=1

\usepackage[T1]{fontenc}
\usepackage{xcolor}
\usepackage{caption, subcaption}
\DeclareCaptionJustification{fulljustify}{%
  \leftskip=0pt\relax
  \rightskip=0pt\relax
  \parfillskip=0pt plus 1fil\relax
  \parindent=0pt\relax
}
\makeatletter
\newcommand{\finishresultfloats}{%
  \par
  \ifx\@deferlist\@empty
    \ifx\@botlist\@empty\else\clearpage\fi
  \else
    \clearpage
  \fi
}
\makeatother
\usepackage{graphicx}
\usepackage{amsmath}
\usepackage{booktabs}
\setcitestyle{sort&compress}

\renewcommand{\,}{\thinspace}

\newcommand{\camb}{\textsc{CAMB}}

\newcommand{\len}{\phi}
\newcommand{\bn}{\hat{\bf n}}

\newcommand{\cmb}{\Theta}

\newcommand{\lm}[1]{l_{#1}m_{#1}} %RAJ
\newcommand{\thrj}[6] { \begin{pmatrix} %RAJ
                       #1 & #2 & #3 \\
		       #4 & #5 & #6
                      \end{pmatrix} }

\usepackage[
  colorlinks=true,
  citecolor=blue,
  linkcolor=blue,
  urlcolor=blue
]{hyperref}

\newlength{\wideeqrulewidth}

\begin{document}

\title{Primordial Power Spectrum from the Lensed CMB Temperature Spectrum using Iterative Delensing and Fourier Denoising}

\author{Rajorshi Sushovan Chandra}
\email[Corresponding author: ]{rajorshi@rrimail.rri.res.in}
% \thanks{Corresponding author.}
\affiliation{Raman Research Institute, C. V. Raman Avenue, Sadashivanagar, Bengaluru 560 080, India}

\author{Debarshi Mukherjee}
\email[Corresponding author: ]{debarshi99mukherjee@gmail.com}
% \thanks{Corresponding author.}
\affiliation{Raman Research Institute, C. V. Raman Avenue, Sadashivanagar, Bengaluru 560 080, India}

\author{Tarun Souradeep}
\affiliation{Raman Research Institute, C. V. Raman Avenue, Sadashivanagar, Bengaluru 560 080, India}
\affiliation{Inter-University Centre for Astronomy and Astrophysics, Pune 411 007, India}

\date{\today}

\begin{abstract}
The CMB power spectra are a strong probe of the primordial power spectrum (PPS) of the 
initial inflationary scalar perturbations. There are many proposed models, from the simpler power law form 
to various well motivated physical models that predict features in the PPS beyond the power law. 
Given the large parameter space of possible models, it is necessary to have 
model independent searches for such features.
Weak gravitational lensing alters the CMB temperature power spectrum and can
bias free form reconstructions of the primordial power spectrum (PPS) if not accounted for precisely, 
such as when using estimators like the Richardson-Lucy (RL) deconvolution algorithm.
We build on our design of the Non linear Iterative Richardson Lucy (NIRL) estimator that
updates the lensing contribution using the freeform PPS recovered at 
the previous iteration, making it truly model independent in the PPS domain while accounting for lensing.
Inspired by signal processing and Fourier filtering techniques from radio interferometry, 
we also design a Fourier space iterative low pass filter 
(FLIP) to suppress noise overfitting in NIRL reconstructions. In conjunction with our previous 
kernel sparsity sampling, we demonstrate the performance of NIRL-FLIP on a range of
simulated PPS models with bootstrap estimates of reconstruction uncertainty obtained from simulated realizations, 
and show that both our estimator and regularization
are robust, accurate and precise, and are able to both reduce lensing induced reconstruction bias
and suppress noise overfitting by a significant factor in Fourier space. 
We provide a detailed description of these results 
and recommend its utility for future model independent searches for hints of novel physics 
in the form of deviations from a power law PPS. This can further
assist in precise and accurate testing of inflation models, in conjunction with joint cosmological parameter estimation.

\end{abstract}

\maketitle

\section{Introduction}
\label{sec:intro}

The inflationary paradigm of cosmological evolution has been exceptionally successful in explaining several 
observed peculiarities in the initial conditions in Cosmology, such as the Horizon problem, the Flatness problem, and the origin of structures. 
Inflation proposes a period of accelerated expansion in the early Universe driven by a scalar inflaton field $\phi$, 
with some potential $V(\phi)$. The potential is usually assumed to be smooth, 
leading to a nearly scale invariant Primordial Power Spectrum (PPS) of the metric perturbations. It is
usually modelled as a power law parametrized by some amplitude $A_s$ and a tilt $n_s$ \cite{Starobinsky:1992ts}.
% [1 A. A. Starobinsky (1992)]
However there are many physically well motivated models that predict features deviating from a power law form.
A non exhaustive list of features includes features that span a wide range of length scales, from large $k$ modes, 
such as a broken spectral index or a step in $n_s$ \cite{JoySahniStarobinsky2008,ContaldiEtAl2003,JainEtAl2009,Stewart1997RunningMass,CoviEtAl2004RunningMass}, 
% [3. Minu Joy, Varun Sahni and Alexei A. Starobinsky (2008), 14. Carlo R. Contaldi, Marco Peloso, Lev Kofman and Andrei Linde (2003),
% [15] Rajeev Kumar Jain, Pravabati Chingangbam, Jinn-Ouk Gong,L. Sriramkumar and Tarun Souradeep (2009), 18a Ewan D. Stewart (1997),
% 18b Laura Covi, David H. Lyth, Alessandro Melchiorri and Carolina J. Odman (2004), ], 
to softer features like localized bumps from field coupling \cite{BarnabyEtAl2009,RomanoSasaki2008}, 
%  [11 Neil Barnaby, Zhiqi Huang, Lev Kofman and Dmitry Pogosyan (2009), [12] Antonio Enea Romano and Misao Sasaki (2008)], 
and different types of oscillatory features, such as from axion monodromy potentials and $H_0$ tension motivated phenomenological fits 
\cite{Hazra:2014goa,AdamsCresswellEasther2001,ChenEastherLim2008,FlaugerEtAl2010,AchucarroEtAl2011,MirandaHuAdshead2012,BeanEtAl2008,Domenech:2019cyh,Beutler:2019ojk}
% [16. Dhiraj Kumar Hazra, Arman Shafieloo, George F. Smoot and Alexei A. Starobinsky (2014), [2] Jennifer A. Adams, Bevan Cresswell and Richard Easther (2001)
% [4] Xingang Chen, Richard Easther and Eugene A. Lim (2008), [5] Raphael Flauger, Liam McAllister, Enrico Pajer, Alexander Westphal and Gang Xu (2010),
% [7] Ana Achucarro, Jinn-Ouk Gong, Sjoerd Hardeman, Gonzalo A. Palma and Subodh P. Patil (2011), [9] Vinicius Miranda, Wayne Hu and Peter Adshead (2012),
% [10] Rachel Bean, Xingang Chen, Girma Hailu, S.-H. Henry Tye and Jiajun Xu (2008), [20] Guillem Domenech and Marc Kamionkowski (2019),
% [S4] Florian Beutler, Matteo Biagetti, Daniel Green, Anze Slosar and Benjamin Wallisch (2019)]
, as well as a combination of these \cite{Polarski:1992dq,JainEtAl2009,AichEtAl2013,GaoLangloisMizuno2012,ChenNamjooWang2015}.
%  [13 David Polarski and Alexei A. Starobinsky (1992), 15. Rajeev Kumar Jain, Pravabati Chingangbam, Jinn-Ouk Gong, L. Sriramkumar and Tarun Souradeep (2009),
% [6] Moumita Aich, Dhiraj Kumar Hazra, L. Sriramkumar and Tarun Souradeep (2013), [8] Xian Gao, David Langlois and Shuntaro Mizuno (2012),
% [17] Xingang Chen, Mohammad Hossein Namjoo and Yi Wang (2015), 
% ]

Since the PPS sources all late time structure formation, it not only helps constrain inflationary models, 
features in PPS could be responsible for  tensions in cosmological 
parameters \cite{Keeley:2020rmo} and CMB anomalies.
The PPS has a direct influence on both the statistics of the CMB
as well as the structure formation and the consequent, weak gravitational lensing of the CMB by the large scale structure of the Universe.
Hence a free form estimation of the PPS and its features provide a direct window into the physics of inflation and the early Universe
\cite{SlosarEtAl2019Features,CoviEtAl2006Step}.
% [Anze Slosar, Xingang Chen, Cora Dvorkin, Daniel Green, P. Daniel Meerburg, Eva Silverstein and Benjamin Wallisch (2019),
% Laura Covi, Jan Hamann, Alessandro Melchiorri, Anze Slosar and Irene Sorbera (2006),
% ]

As experiments have improved in precision and accuracy in successive results from WMAP, Planck, SPT and ACT, 
reconstructions of the PPS have narrowed closer and closer
towards a phenomenological power law form close to the idealised classic scale-invariant Harrison-Zeldovich spectrum  
\cite{HinshawEtAl2013WMAP9,Aghanim:2018eyx,Aghanim:2019ame,Akrami:2018odb,Rosenberg:2022sdy}.
% [[1] G. Hinshaw et al. (2013), [2] N. Aghanim et al. [Planck] (2020), [3] Y. Akrami et al. [Planck] (2020), 
% [4] E. Rosenberg, S. Gratton and G. Efstathiou (2022)]
However, with increasing precision of data, and larger $l$ modes being probed, especially by the recent SPT and ACT observations
\cite{LouisEtAl2025ACTDR6,CamphuisEtAl2026SPT3GD1},
% [[5] T. Louis et al. [Atacama Cosmology Telescope] (2025), [6] E. Camphuis et al. [SPT-3G] (2026)],
there continues to remain room for further investigation of the PPS, especially in the context of anomalies and tensions
like the Hubble tension. In addition, hitherto noise dominated, high $l$ regions opening up leads
to novel avenues of designing unbiased estimators, such as accounting for lensing.
Departures and joint constraints of anomalies have been explored previously, such as through Wiggly Whipped Inflation
\cite{Hazra:2014goa,HazraEtAl2016PlanckPolarization}
%  [[7] D. K. Hazra, A. Shafieloo, G. F. Smoot and A. A. Starobinsky (2014),
% [8] D. K. Hazra, A. Shafieloo, G. F. Smoot and A. A. Starobinsky (2016)]
, though further analysis from Planck 2018 and SPT data shows either a preference for the smooth slow-roll form 
or possible features as not being statistically significant
\cite{HazraEtAl2021InflationStory,MeerburgEtAl2014PlanckOscillations,AntonyEtAl2026PlanckSPTOscillations}.
%  [[9] D. K. Hazra, A. Shafieloo, G. F. Smoot and A. A. Starobinsky (2021), [10] P. D. Meerburg, D. N. Spergel and B. D. Wandelt (2014),
% [11] A. Antony, F. Finelli, D. K. Hazra, D. Paoletti and A. Shafieloo (2026)].

Reconstruction methods fall broadly into two categories, parametric model dependent inference and model independent free form reconstruction.
Much of the parametric forms follow from the earlier mentioned departure from power law forms, where the extra model based
parameters are fit jointly with the standard cosmological parameters. There are hybrid methods, such as Bayesian knot inference 
where the knot parameters are not physically motivated from a particular model, but slide around to best converge to the data and hint
at departures from the power law, and other methods like smoothed cubic splines.
\cite{RaffaelliBallardini2025Knots,BridgesEtAl2009PPS,AslanyanEtAl2014Knots,HandleyEtAl2019Reconstruction,SealfonEtAl2005Splines,VerdePeiris2008Splines}.
%  [[18] A. Raffaelli and M. Ballardini (2025), [2] M. Bridges, F. Feroz, M. P. Hobson and A. N. Lasenby (2009),
% [3] G. Aslanyan, L. C. Price, K. N. Abazajian and R. Easther (2014), [4] W. J. Handley, A. N. Lasenby, H. V. Peiris and M. P. Hobson (2019),
% [5] A. Raffaelli and M. Ballardini (2025), [6] C. Sealfon, L. Verde and R. Jimenez (2005), [7] L. Verde and H. V. Peiris (2008)
% ].
There are direct inversion techniques like likelihood inversion and Tikhonov regularization
\cite{TocchiniEtAl2005PPS,HuntSarkar2014Tikhonov,HuntSarkar2015Planck,GauthierBucher2012PPS,NagataYokoyama2009Bandpowers},
%  [[8] D. Tocchini-Valentini, M. Douspis and J. Silk (2005), [9] P. Hunt and S. Sarkar (2014),
% [10] P. Hunt and S. Sarkar (2015),[11] C. Gauthier and M. Bucher (2012), [12] R. Nagata and J. Yokoyama (2009)].
as well as cosmic inversion methods 
\cite{MatsumiyaEtAl2002Inversion,MatsumiyaEtAl2003Inversion,KogoEtAl2004Inversion},
%  [[13] M. Matsumiya, M. Sasaki and J. Yokoyama (2002), [14] M. Matsumiya, M. Sasaki and J. Yokoyama (2003),
% [15] N. Kogo, M. Matsumiya, M. Sasaki and J. Yokoyama (2004)]
singular value decomposition and principal component analysis methods as an alternative linear algebra fit to the PPS 
\cite{2010JCAP...01..016N,Leach2006PCA},
%  [[16] G. Nicholson, C. R. Contaldi and P. Paykari (2010), [17] S. M. Leach (2006)]
and wavelet basis inversion and fitting, such as the PRISM algorithm
\cite{MukherjeeWang2003Wavelets,2014A&A...566A..77P,Lanusse:2014sra}.
% [[18] P. Mukherjee and Y. Wang (2003), [19] P. Paykari, F. Lanusse, J.-L. Starck, F. Sureau and J. Bobin (2014),
% [20] F. Lanusse, P. Paykari, J.-L. Starck, F. Sureau, J. Bobin and A. Rassat (2014),].
There have been other novel methods tested recently, such as Gaussian processes combined with neural networks, and genetic algorithms 
built from a function database to evolve to analytic PPS expressions
\cite{ChakiEtAl2025GP,LodhaEtAl2024GA}.
% [[22] S. Chaki, A. Nicola, A. Spurio Mancini, D. Piras and R. Reischke (2025), 
% [23] K. Lodha, L. Pinol, S. Nesseris, A. Shafieloo, W. Sohn and M. Fasiello (2024)]

In this work we look at a specific class of free form reconstruction methods, known as the Richardson-Lucy deconvolution algorithm.
The fundamental premise of this iterative deconvolution method is that it is a gradient descent estimator of an underlying input field
in the presence of a convolution kernel, and it has certain challenges while dealing with noise, applied in both astronomy and optical imaging 
such as photography and microscopy \cite{1972JOSA...62...55R,1974AJ.....79..745L,1992AJ....104.1260L,1994A&A...289..983L,DeyEtAl2006RLTV,SchmittEtAl2012PoissonSphere}.
% [W. H. Richardson. "Bayesian-Based Iterative Method of Image Restoration." J. Opt. Soc. Am. 62, 55--59 (1972).
% L. B. Lucy. "An iterative technique for the rectification of observed distributions." Astron. J. 79, 745--754 (1974),
% L. B. Lucy. Astron. J. 104, 1260--1265 (1992),
% L. B. Lucy. Astron. Astrophys. 289, 983--994 (1994).
% N. Dey, L. Blanc-Feraud, C. Zimmer, P. Roux, Z. Kam, J.-C. Olivo-Marin and J. Zerubia. Microsc. Res. Tech. 69, 260--266 (2006),
% J. Schmitt, J.-L. Starck, J.-M. Casandjian, J. Fadili and I. Grenier. Astron. Astrophys. 546, A114 (2012),
% ]
There have been various modifications of this method to incorporate noise regularization
and adapt the estimator to the properties of the convolution kernel. Originally it was used for image deconvolution from the point spread 
function. In the case of cosmology, it has been applied to power spectrum deconvolution
\cite{1993MNRAS.265..145B,1994MNRAS.267..323B}.
% [C. M. Baugh and G. Efstathiou. Mon. Not. R. Astron. Soc. 265, 145--156 (1993).
% C. M. Baugh and G. Efstathiou. Mon. Not. R. Astron. Soc. 267, 323 (1994).]
Specifically for CMB, the radiative transport kernel behaves as the convolving function and the underlying $P(k)$ is the
underlying observable to be deconvolved, in the presence of observation noise of the CMB temperature and polarization power spectra. 
The RL estimator has a feature where the convolution quantities, including the kernel and the input plus output vectors, are all positive definite everywhere
by construction, a consequence of the maximum likelihood estimator representation of a Poissonian distribution \cite{SheppVardi1982MLE}. 
This is a useful property for our use case, where our cosmological quantities of interest are positive definite everywhere.
Several works on this topic exist, with many of them adding regularization at different stages, such as noise weighting of the input data
and noise regularization of the PPS reconstruction through variance and gradient control or wavelet decomposition
\cite{Shafieloo:2003gf,Shafieloo:2006hs,Nicholson:2009pi,SohnEtAl2022PPS}.

% \cite{SheppVardi1982MLE}

% [A. Shafieloo and T. Souradeep. Phys. Rev. D 70, 043523 (2004).
% A. Shafieloo, T. Souradeep, P. Manimaran, P. K. Panigrahi and R. Rangarajan. Phys. Rev. D 75, 123502 (2007),
% G. Nicholson and C. R. Contaldi. JCAP 07, 011 (2009),
% W. Sohn, A. Shafieloo and D. K. Hazra. JCAP 03, 056 (2024).
% ]

The RL method and its variations like the Modified Richardson-Lucy (MRL) algorithm with Gaussian smoothing, $tanh$ noise weighting
have been applied to successive CMB data releases to explore deviations from a power law PPS, and also improve joint fitting of
background cosmology parameters and initial condition parameters together
\cite{Shafieloo:2007tk,2010JCAP...04..010H,2011NJPh...13j3024S,Hazra:2013xva,Hazra:2013eva,HazraEtAl2024PlanckACT,Shaikh:2016hrf}. 
% [A. Shafieloo and T. Souradeep. Phys. Rev. D 78, 023511 (2008),
% J. Hamann, A. Shafieloo and T. Souradeep. JCAP 04, 010 (2010),
% A. Shafieloo and T. Souradeep. New J. Phys. 13, 103024 (2011),
% D. K. Hazra, A. Shafieloo and T. Souradeep. JCAP 07, 031 (2013),
% D. K. Hazra, A. Shafieloo and T. Souradeep. Phys. Rev. D 87, 123528 (2013),
% D. K. Hazra, B. Beringue, J. Errard, A. Shafieloo and G. F. Smoot. JCAP 12, 038 (2024),
% ]
The most recent updates to this include a new 
heat diffusion based regularization of the PPS, applied to a three data set combination with different late time cosmologies, 
with results showing overall consistency of the data with a power law PPS. While there are local departures from power law in
individual data sets of SPT and ACT, they do not show any evidence of significant cross data set features which are statistically significant
\cite{SohnEtAl2022PPS,ChandraEtAl2026PPS}.
% [W. Sohn, A. Shafieloo and D. K. Hazra. JCAP 03, 056 (2024),
% D. Chandra, D. K. Hazra, A. Shafieloo and T. Souradeep. arXiv preprint, version 1 (2026).
% ]

Since there is now evidence of degeneracies between the early universe PPS feature space and late time cosmological parameters, 
it is now imperative to have a more accurate and robust understanding of reconstruction systematics that dominate the $k$ modes 
of maximum reconstruction sensitivity, such as CMB weak lensing \cite{Hazra:2014jwa,Hazra:2018opk,Keeley:2020rmo,HazraAntonyShafieloo2022OneSpectrum}. 

In this work we quantify the weak-lensing contribution to the temperature
transport kernel and test an iterative PPS-independent correction within
the MRL reconstruction. We use feature injections to quantify the new
iterative lensing correction and its convergence with respect to analysis 
parameters like iterations and noise injections modeled after expected real data.
We also use the Fourier structure of the noisy reconstructions,
motivated by the information-based analysis of RL noise amplification in
Ref.~\cite{RLnature2025}, to design an iterative low-pass regularizer.
Together with sparse kernel sampling \cite{Chandra:2021ydm}, these steps
reduce reconstruction error in the simulations and motivate convergence
diagnostics calibrated to the noise and retained feature scales.

\section{Weak lensing correction to the temperature power spectrum}
\label{sec:cltt_lens_harmonic}

\begin{figure*}[!tbp]
\centering
\begin{subfigure}[t]{0.49\textwidth}
  \includegraphics[width=\linewidth]{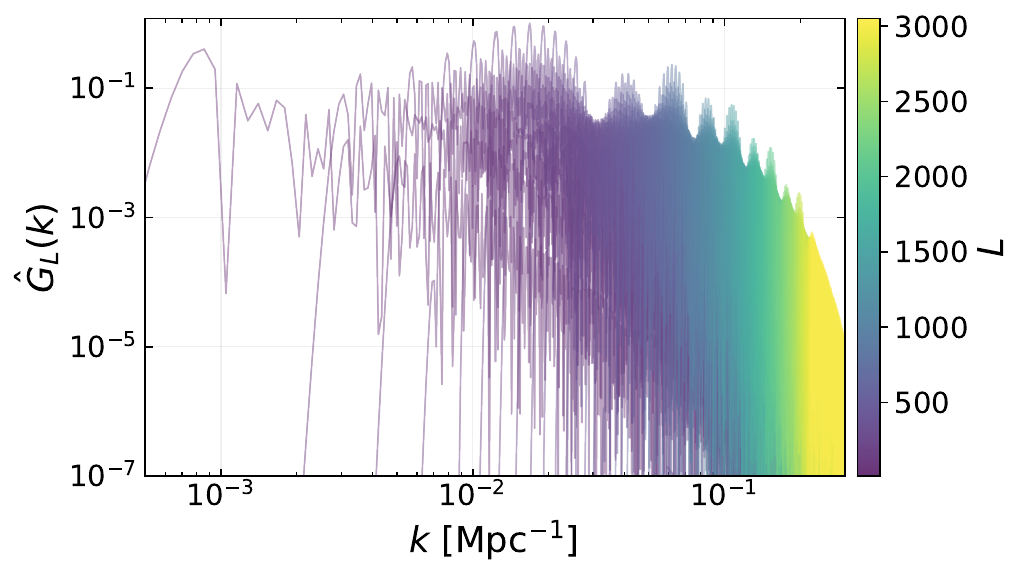}
  \caption{}
  \label{fig:ft_glk_production}
\end{subfigure}\hfill
\begin{subfigure}[t]{0.49\textwidth}
  \includegraphics[width=\linewidth]{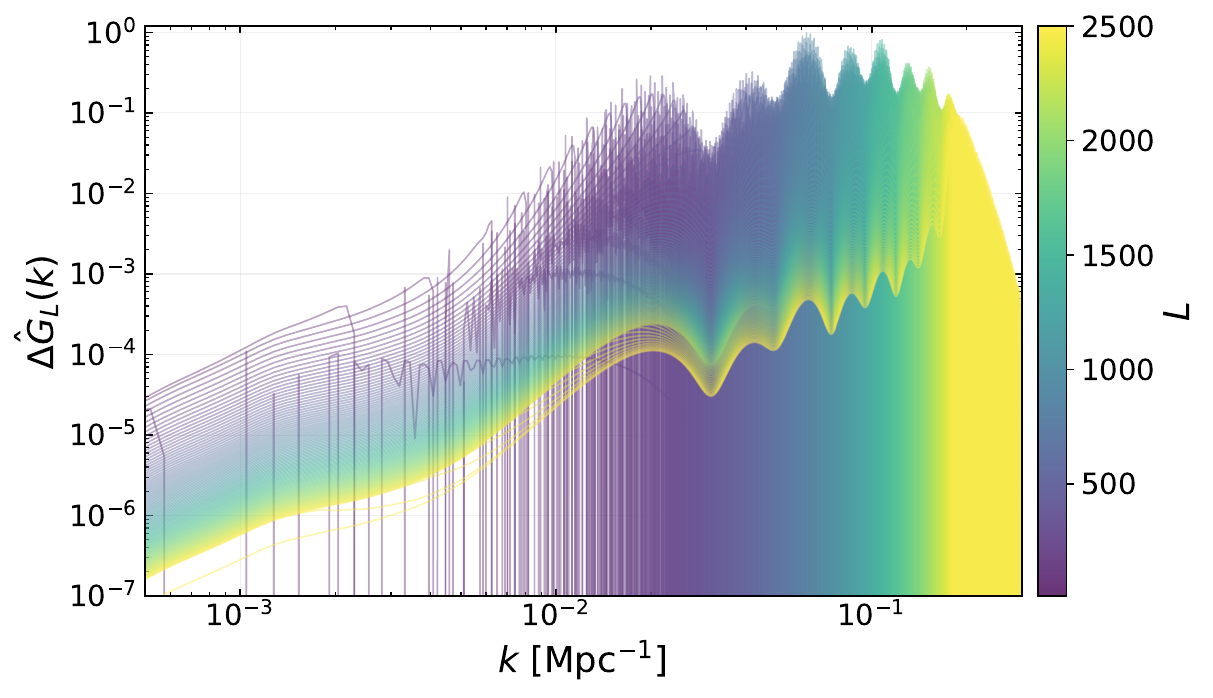}
  \caption{}
  \label{fig:glk_production}
\end{subfigure}
\caption{Peak normalized temperature transport kernel
$\hat G_L(k)$, with selected multipoles $L$ indicated by colour for unlensed (a) and lensing correction (b) kernels. 
The kernel weight varies with $L$ and $k$ showing how a model independent freeform reconstruction is sensitive to
weak lensing at high $k$ regions.}
\label{fig:glk_ft_pair}
\end{figure*}

In weak gravitational lensing of the CMB, the CMB photons from the last scattering surface to the observer, are deflected by
the intervening large scale structure of the Universe. This can be understood under Born approximation as a remapping of the CMB temperature field,
by a vector which is the transverse gradient of the projected lensing potential, with respect to the line of sight direction\cite{Lewis:2006fu}.

% remapping Here
\begin{equation}
\widetilde{\Theta}(\bn)
=
\Theta\!\left[\bn+\boldsymbol{\alpha}(\bn)\right] \,,
\qquad
\alpha_i(\bn)=\nabla_i\phi(\bn) \,.
\label{eqn:lensing_remapping}
\end{equation}
$\nabla_i$ is the covariant angular derivative on the unit sphere, hence
$\boldsymbol{\alpha}$ is the two-dimensional deflection field transverse to $\bn$.

Following the harmonic-space formalism of \cite{Hu:2000ee}, the lensed temperature multipoles expanded to second order in
the lensing potential $\phi_{LM}$ read
\begin{equation}
\begin{split}
\widetilde{\Theta}_{{\ell}m} &= {\Theta}_{{\ell}m}
+ \sum_{\lm{1}} \sum_{\lm{2}} {\Theta}_{\lm{2}} \phi_{\lm{1}} I_{{\ell} l_1 l_2}^{m m_1 m_2} \\
&\quad + {1 \over 2} \sum_{\substack{l_1m_1 \\ l_2m_2}} \sum_{l_3 m_3} \len_{l_1 m_1} \len_{l_3 m_3}^* \cmb_{l_2
m_2} J_{{\ell} l_1 l_2 l_3}^{m m_1 m_2 m_3} \,,
\end{split}
\label{eqn:delens_start1}
\end{equation}
with geometric shape factors
\begin{equation}
\begin{split}
I_{{\ell} l_1 l_2}^{m m_1 m_2} & = \int d\bn\, Y_{\ell m}^*
( \nabla_i Y_{l_1 m_1} ) ( \nabla^i Y_{l_2 m_2} ) \,, \\
J_{{\ell} l_1 l_2 l_3}^{m m_1 m_2 m_3} & = \int d\bn\, Y_{\ell m}^*
( \nabla_i Y_{l_1 m_1} ) ( \nabla_j Y_{l_3 m_3}^* )\, \nabla^i \nabla^j Y_{l_2 m_2} \,.
\end{split}
\end{equation}
Assuming Gaussianity and statistical isotropy of both fields, the two-point function gives the lensed spectrum
to first order in $C_L^{\phi\phi}$,
\begin{equation}
\begin{split}
\widetilde{C}_{\ell}^{TT}
 &= {C}_{\ell}^{TT} + {C}_{\ell}^{TT} \sum_{l_1}
 {C}_{l_1}^{\phi\phi} S^{(b)}_{{\ell}l_1} \\
 &\quad + \sum_{l_1l_2} {C}_{l_1}^{\phi\phi}
 {C}_{l_2}^{TT} S^{(a)}_{{\ell}l_1l_2} \,,
\end{split}
\label{eqn:lencltt_sum}
\end{equation}
where standard spherical-harmonic identities \cite{Hu:2000ee,Lewis:2006fu,2011PhRvD..83d3005H} reduce the shape-factor sums to
\begin{equation}
\begin{split}
S^{(a)}_{{\ell}l_1l_2} &= \frac{1}{2{\ell}+1} (F_{{\ell}l_1l_2})^2 \,, \\
S^{(b)}_{{\ell}l_1} &= -\frac{1}{2} {\ell}({\ell}+1)\, l_1(l_1+1)\, \frac{2l_1+1}{4\pi} \,, \\
\widetilde{C}_{\ell}^{TT} &= {C}_{\ell}^{TT} + \Delta{C}_{\ell}^{(b)TT} + \Delta{C}_{\ell}^{(a)TT} \,,
\end{split}
\label{eqn:lencltt_deriv}
\end{equation}
with
\vspace{-0.6\baselineskip}
\begin{equation}
\begin{split}
F_{l_1l_2l_3} = \frac{1}{2} & [l_2(l_2+1)+l_3(l_3+1)-l_1(l_1+1)] \\
& \hspace{-4em}\times \sqrt{\frac{(2l_1+1)(2l_2+1)(2l_3+1)}{4\pi}}
\thrj{l_1}{l_2}{l_3}{0}{0}{0} \,.
\end{split}
\label{Fll'Leven}
\end{equation}
Equation~\eqref{eqn:lencltt_deriv} is the familiar peak-smoothing of the temperature spectrum, written as an explicit operator acting
on $\{C_\ell^{TT}, C_L^{\phi\phi}\}$. Its importance here is an algorithmic property arising from physics; both spectra on the right-hand side are
convolutions of the \emph{same} $P_R(k)$, which is what renders the lensed spectrum a non-linear functional
of the PPS and motivates the estimator of the next section.

\section{Method: non-linear iterative Richardson--Lucy deconvolution}
\label{sec:NIRLdeconv}

\subsection{The non-linear convolution}

The RL algorithm and its cosmological variants \cite{1972JOSA...62...55R,1974AJ.....79..745L,1993MNRAS.265..145B,1994MNRAS.267..323B,Shafieloo:2003gf,Hazra:2014jwa,Chandra:2021ydm} 
operate on convolutions represented numerically in the form,
\begin{equation}
C_{\ell}^{XX} = \sum_{k} P_k\, G_{{\ell}k}^{XX} \,,
\end{equation}
where $G_{\ell k}^{XX}$ is the discretised radiative transport kernel. Writing both the unlensed temperature and
lensing-potential spectra in this form,
\begin{equation}
{C}_{\ell}^{TT} = \sum_{k} P_{k}\, G_{{\ell}k}^{TT} \,, \qquad
{C}_L^{\phi\phi} = \sum_{k_1} P_{k_1}\, G_{Lk_1}^{\phi\phi} \,,
\end{equation}
If we now substitute the unlensed $C_{\ell}^{TT}$ and $C_L^{\phi\phi}$ into equation ~\eqref{eqn:lencltt_sum}
, we end up with the composite lensed temperature spectrum as a second order function of the PPS 
where the effective kernel $\tilde{G}_{\ell k}^{TT}$ depends on $P_k$: 

\setlength{\wideeqrulewidth}{\columnwidth}
\onecolumngrid
\vspace{1.5\baselineskip}
\noindent\makebox[\textwidth][l]{%
  \rule{\wideeqrulewidth}{0.4pt}\rule{0.4pt}{6pt}%
}\par
\begin{equation}
\tilde{C}_{\ell}^{TT} = \sum_{k} P_{k}\, \tilde{G}_{{\ell}k}^{TT}
 = \sum_{k} P_{k}
\bigg[
G_{{\ell}k}^{TT} \bigg(1 + \sum_{k_1} P_{k_1} \sum_{l_1} G_{l_1k_1}^{\phi\phi} S^{(b)}_{{\ell}l_1} \bigg) + \bigg( \sum_{k_1} P_{k_1} \sum_{l_1l_2} G_{l_1k_1}^{\phi\phi}
G_{l_2k}^{TT} S^{(a)}_{{\ell}l_1l_2} \bigg)
\bigg] \,.
\label{eqn:nonlinconv}
\end{equation}
\noindent\makebox[\textwidth][r]{%
  \rule[-6pt]{0.4pt}{6pt}\rule{\wideeqrulewidth}{0.4pt}%
}\par
\twocolumngrid

\begin{figure}[!tbp]
\includegraphics[width=1.00\linewidth]{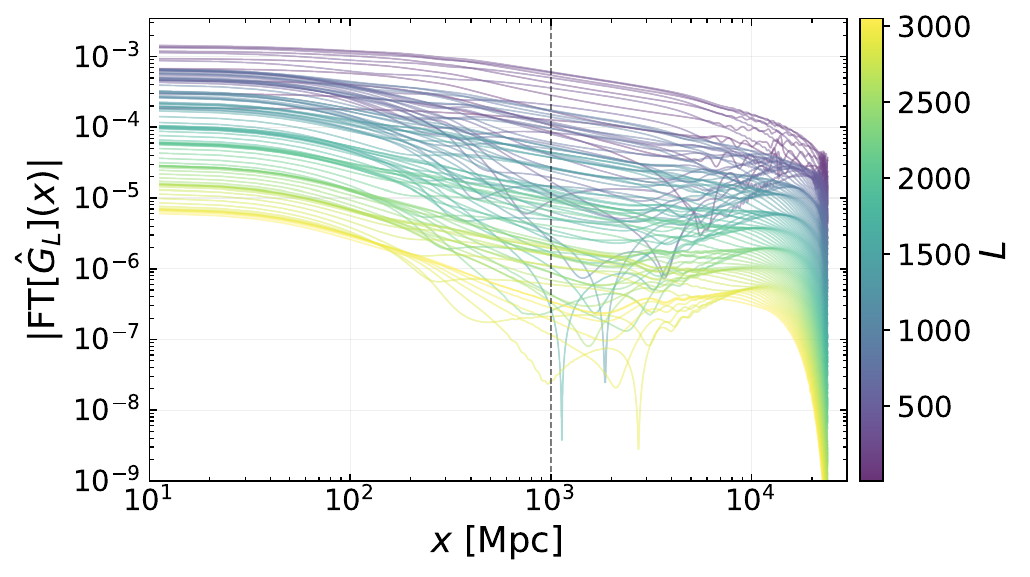}
\caption{Fourier amplitudes of the peak normalized radiative transport kernel,
with select multipole $L$ indicated by colour. $\xi$ is the conjugate coordinate to $k$. 
The dashed line at $10^3$\,Mpc marks the filter cutoff used in the regularization filter.}
\label{fig:ft_glk}
\end{figure}

\begin{figure}[!tbp]
\includegraphics[width=1.00\linewidth]{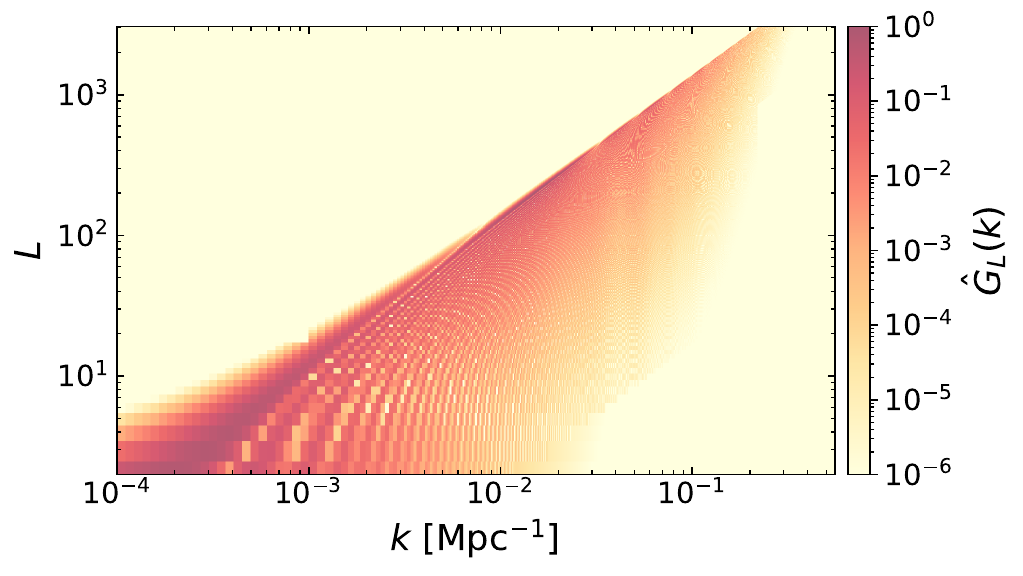}
\caption{Peak normalized unlensed temperature transport kernel in the $(L,k)$ plane,
shown on a logarithmic colour scale. The broad tails below the principal
ridge depict the coupling of $k$ mode to multiple temperature
multipoles, but broadly exhibit a diagonal sensitivity.}
\label{fig:glk_heatmap}
\end{figure}
The physical intuition of this equation is a visualization of mode mixing, where lensing distributes power over different length scales 
across the sky. This leads to a smoothing of the acoustic peaks in the power spectrum and fills out the troughs. Due to Silk
damping, at high $\ell$ modes, lensing contribution to the power becomes more significant and dominant. 
This is also the region where cosmic variance is lower and high resolution instruments are capable of measuring 
the power spectrum with high precision, though challenges from foregrounds and mask effects remain.
This naturally means that while searching for features in the PPS from power spectrum data, the lensing contribution 
behaves akin to a systematic that needs to be accounted for an unbiased estimator of the PPS. 
If a direct RL deconvolution is applied on the lensed temperature spectrum, the lensing smoothing contribution is 
reconstructed as spurious oscillatory departures from a power law PPS, in the ranges of  
$k \sim 3\times10^{-2}$--$2\times10^{-1}\,\mathrm{Mpc}^{-1}$. 
(A trace of this is visible in row 3 of Fig.~\ref{fig:pkrecon_all}, as the blue spikes in template delensing, which leaks 
significant lensing power.)
These spurious features can further percolate into cosmology parameters if jointly fit.
A first order correction in the case of RL deconvolution, which works well for reasonably noisy data, 
is to subtract a lensing template, where the template is usually constructed from fiducial late time cosmology 
and a power law PPS ansatz \cite{Hazra:2014jwa}. This works reasonably well in the presence of sufficiently noisy data
with different regularization schemes.

\subsection{The NIRL estimator}

However, as the noise floor deepens and approaches the limit of cosmic variance at high $\ell$ modes, the propagated error bars on a 
PPS reconstruction shrink drastically. This means any departures from power law become statistically significant. 
Since the goal of the RL deconvolution is to reconstruct the PPS without any model assumptions, the template approach in 
principle is not ideal since it assumes a power law PPS to construct the lensing template. This implies that should
the true PPS deviate beyond the power law, the template will not account for some portion of the lensing contribution, 
and the RL deconvolution remains biased.
We therefore introduce the Nonlinear Iterative Richardson Lucy (NIRL) estimator where, under the physically motivated condition
that the departures from power law are a fraction of total PPS power, we iteratively add the kernel lensing contribution 
calculated at each iteration step using the current PPS reconstruction, to the deconvolution transport kernel until convergence.
This ensures that our NIRL algorithm starts from a true model independent PPS starting guess and iteratively corrects the 
lensing systematic in situ inside each NIRL iteration step. The NIRL estimator is now a more unbiased estimator  
and model independent with respect to early Universe physics. In the absence of lensing NIRL reduces to the standard MRL estimator.
Conversely, any early universe physics sourced secondary anisotropies similar to lensing can be incorporated into the NIRL estimator
with the appropriate correction kernel update.

At iteration $i$, the kernel is corrected using the current estimate $P_k^{(i)}$,
\begin{equation}
\begin{split}
{\tilde{G}}_{\ell k}^{TT (i)} &=
G_{\ell k}^{TT} \bigg[1 + \sum_{k_1} P_{k_1}^{(i)} \sum_{l_1} G_{l_1k_1}^{\phi\phi} S^{(b)}_{\ell l_1} \bigg] \\
&\quad + \bigg[ \sum_{k_1} P_{k_1}^{(i)} \sum_{l_1l_2} G_{l_1k_1}^{\phi\phi} G_{l_2k}^{TT} S^{(a)}_{\ell l_1l_2} \bigg] \\
&= G_{\ell k}^{TT} + \Delta G_{\ell k}^{TT}\big(P^{(i)}\big) \,,
\end{split}
\label{eqn:numer_lenscorr}
\end{equation}
and the update inherits the multiplicative RL form,
\begin{equation}
\begin{split}
P_{k}^{(i+1)} &= P_{k}^{(i)} \bigg[ 1 +
\sum_{\ell} \hat{\tilde{G}}_{\ell k}^{TT (i)} \bigg( \frac{\hat{C}_{\ell}^{TT}}{C_{\ell}^{TT(i)}} - 1 \bigg) \bigg] \,, \\
C_{\ell}^{TT(i)} &=  \sum_{k} {\tilde{G}}_{\ell k}^{TT (i)} P^{(i)}_k \,,
\end{split}
\label{eq:IRL_eqns_nonlin_notanerr}
\end{equation}
where $\hat{C}_{\ell}^{TT}$ is the (lensed) data and $\hat{\tilde{G}}_{\ell k}^{TT(i)}$ is normalised over $\ell$. 
In this approach we correct the kernel instead of delensing the data with an iteration-built template.

For noisy data we combine the error-weighted MRL form \cite{1993MNRAS.265..145B,1994MNRAS.267..323B,Shafieloo:2003gf,Chandra:2021ydm}, given by
\begin{equation}
\begin{split}
P_{k}^{(i+1)} &= P_{k}^{(i)} \bigg[ 1 +
\sum_{\ell} \hat{\tilde{G}}_{\ell k}^{TT (i)}
\bigg( \frac{\hat{C}_{\ell}^{TT}}{C_{\ell}^{TT(i)}} - 1 \bigg) \\
&\qquad\quad \times \text{tanh}^2
\bigg( \frac{\hat{C}_{\ell}^{TT}-C_{\ell}^{TT(i)}}{\hat{\sigma}_{\ell}} \bigg)^2 \bigg] \,,
\end{split}
\label{eq:IRL_eqns}
\end{equation}
with $\hat{\sigma}_\ell$ the measurement error on $\hat{C}_{\ell}^{TT}$, 
and $\hat{\sigma}_\ell^2$ is the diagonal covariance element. The expression can be rewritten for a complete covariance matrix
in the general case.
The convergence criteria for NIRL in the case of noisy data, remains similar to previous methods, 
where we use the $\chi^2_{(i)}$ statistic at each iteration $i$ to quantify the goodness of fit of the 
current PPS-to-power spectrum estimate versus the data, given by
\begin{equation}
\chi^2_{(i)} = \sum_\ell \frac{\big(\hat{C}_\ell^{TT} - C_\ell^{TT(i)}\big)^2}{\hat{\sigma}_\ell^2} \,,
\label{eq:chi2def}
\end{equation}
For noisy data there are no definitive stopping criteria for the RL algorithm.
We base our stopping criteria on the $\chi^2$ statistic, 
where we iterate to a sufficiently high number of iterations,
and we select the iteration with the minimum $\chi^2$ as the best fit PPS.
Since any RL estimator is prone to noise overfitting, which gets worse when features of interest
occupy similar $k$ scales to the noise, we defer regularization and noise control to Sec.~\ref{sec:regularization}.

\section{Transport kernel: simulation, precision, and sparse grid}
\label{sec:cltt_kernelsim}

\begin{figure}[!tbp]
\includegraphics[width=1.00\linewidth]{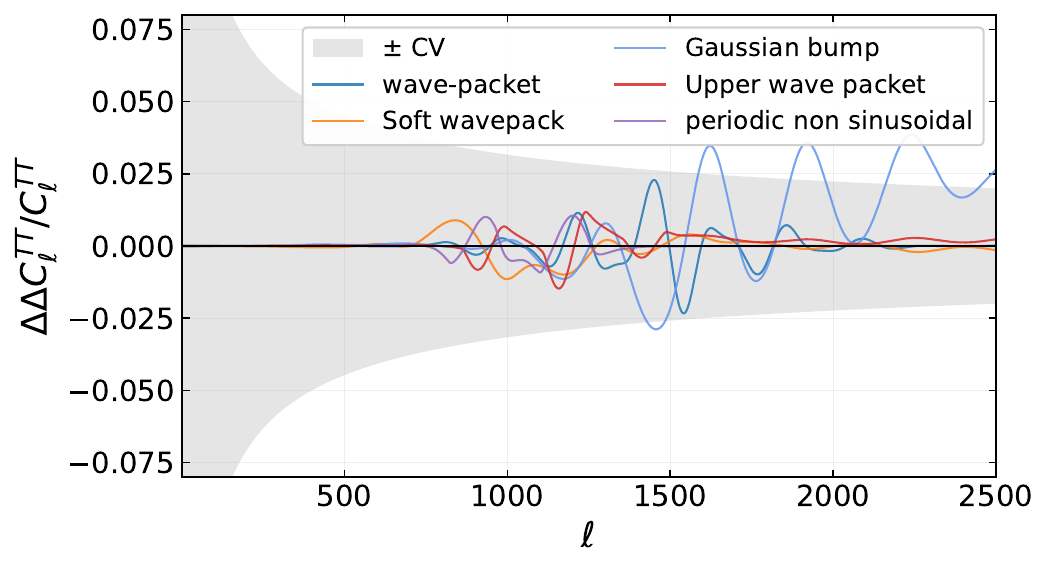}
\caption{Relative difference between the simulated feature and power law lensing corrections,
$\Delta\Delta C_\ell^{TT}/C_\ell^{TT}
=(\Delta C_\ell^{\rm feat}-\Delta C_\ell^{\rm pl})/C_\ell^{\rm ref}$,
for the five features in Fig.~\ref{fig:pkinput_all}. The grey envelope is
$\pm\sqrt{2/(2\ell+1)}$.}
\label{fig:features_overlay}
\end{figure}

\begin{figure}[!tbp]
\includegraphics[width=1.00\linewidth]{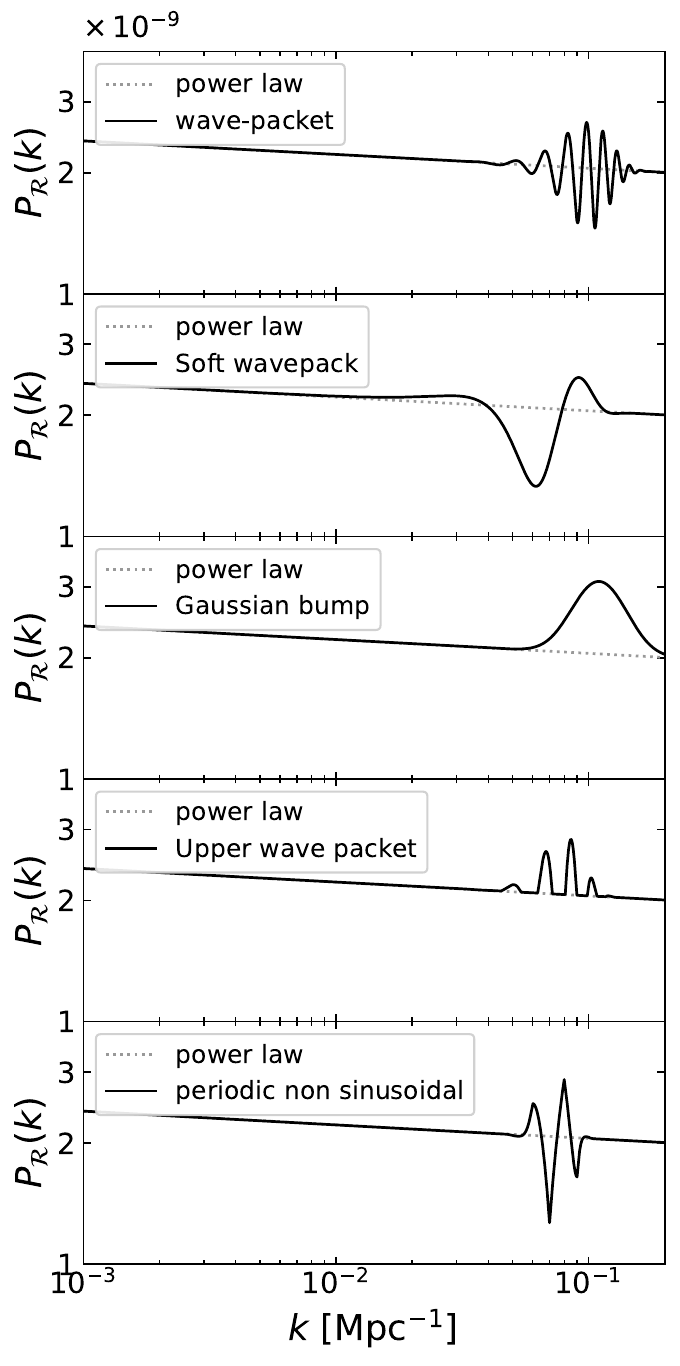}
\caption{Input test PPS for five features spanning varying $k$ ranges and resolutions, 
with the smooth power law shown in dotted lines as a reference.}
\label{fig:pkinput_all}
\end{figure}

\begin{figure}[!tbp]
\includegraphics[width=1.00\linewidth]{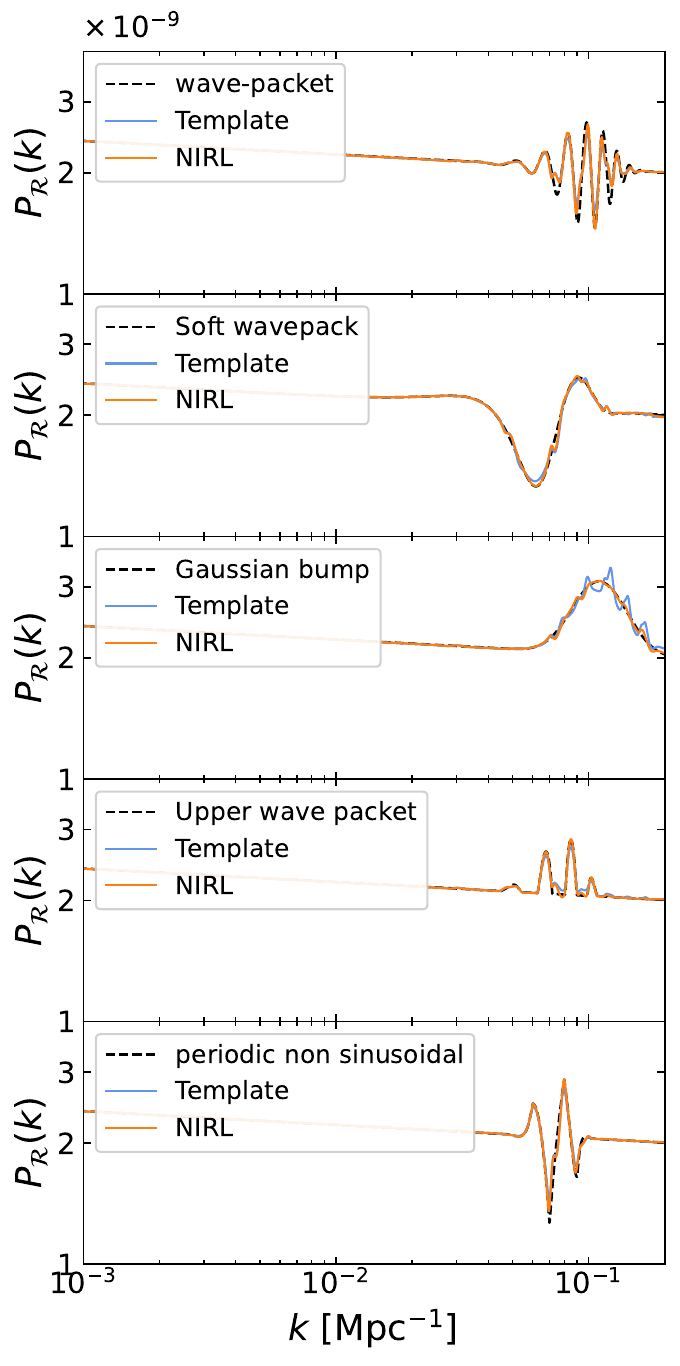}
\caption{Reconstructions for the five input PPS in
Fig.~\ref{fig:pkinput_all} for 500 fixed iterations (Without noise realisation, there is no $\chi^2$ fit comparison). 
Template (blue) and NIRL (orange) are overlaid on the injected truth (dashed).}
\label{fig:pkrecon_all}
\end{figure}

\begin{figure}[!tbp]
\includegraphics[width=1.00\linewidth]{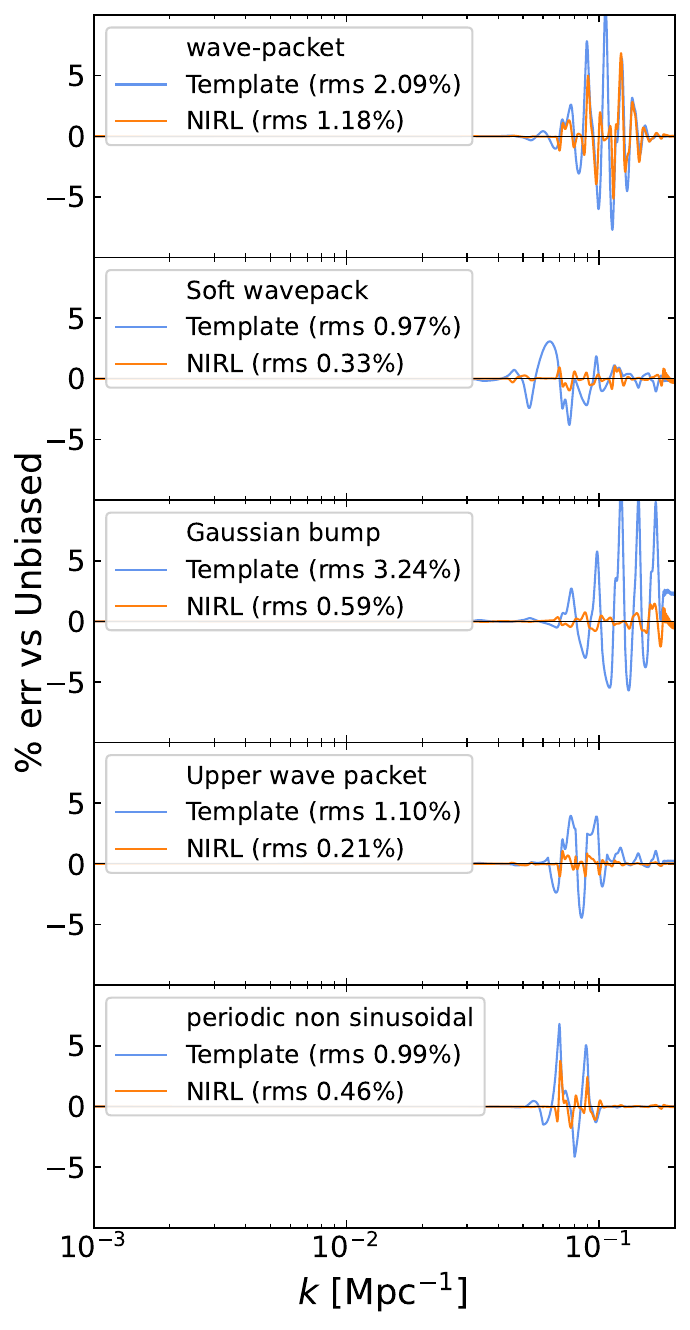}
\caption{Reconstruction relative error plots from each feature for template (blue) versus NIRL (orange), 
with respect to the unbiased reconstruction. Legend values give the average over $k$ rms over
$0.02<k<0.3\,\mathrm{Mpc}^{-1}$ [Eq.~\eqref{eq:feature_rms}], tabulated in
Table~\ref{table:battery}.}
\label{fig:pkreconerr_all}
\end{figure}

\subsection{Kernel simulation}
\label{sub_sec:kernel_behav_lens}

We generate the temperature and lensing-potential transport kernels,
$G_{\ell k}^{TT}$ and $G_{Lk}^{\phi\phi}$, with \camb{} \cite{Lewis:1999bs},
using the Planck best-fit $\Lambda$CDM parameters in Table~\ref{table:phy_param}
\cite{Aghanim:2018eyx} and enhanced sampling (\texttt{lSampleBoost}=51).
Figure~\ref{fig:glk_ft_pair} shows the peak normalised unlensed temperature kernel $G_{\ell k}^{TT}$
as a function of $k$ for selected multipoles $L$ in part (a) 
and the equivalent peak normalised lensing contribution $\Delta G_{\ell k}^{TT}$ in part (b). 
The acoustic structure is clearly visible as well as the dominance of lensing power at high multipoles
which strongly motivates our in situ NIRL estimator. 

The main precomputation for Eq.~\eqref{eqn:numer_lenscorr} is the three-index
shape factor $S^{(a)}_{\ell l_1l_2}$. Its stored representation occupies
approximately 30\,GB, after which each NIRL kernel update takes approximately
2 minutes in our implementation on any standard laptop CPU such as the Intel i5 8th generation and above. 
We use $l_1\in[2,2500]$ and
$l_2\in[2,3000]$ for reconstructions through $\ell_{\max}=2500$. We also extend the
temperature summation to an $l_{2,\max}=\ell_{\max}+500$ so that it includes coupling to
multipoles beyond the reconstruction range. This essentially provides a buffer range for 
high accuracy mode mixing.

\begin{table}[htb]
\begin{center}
\begin{tabular}{ l r }
  \toprule
  Parameter & Planck best fit \\
  \midrule
  $H_0$ & 67.36 \\
  $\Omega_b h^2$ & 0.02237 \\
  $\Omega_c h^2$ & 0.1200 \\
  $m_\nu$ & 0.06 eV \\
  $\Omega_K$ & 0 \\
  $\tau$ & 0.0544 \\
  \bottomrule
\end{tabular}
\end{center}
\caption{Cosmological parameters \cite{Aghanim:2018eyx} used for the \camb{} radiative transport kernels.}
\label{table:phy_param}
\end{table}

\subsection{On $k$ grid Kernel Sparsity Sampling}
\label{sec:sparse}

We use the Kernel Sparsity Sampling (KSS) from our previous work, 
which reduces the number of PPS degrees of freedom to stay within the number of data points, 
while maximizing the kernel response \cite{Chandra:2021ydm}. Starting from the dense $k$ grid, the
algorithm identifies segments according to the number of multipoles with
peak kernel response per $L$ at each $k$. Within a segment, it compares the
multipole count, $\Delta N_L$, with the available
dense-grid count, $\Delta N_k$, and retains
$\min(\Delta N_L,\Delta N_k)$ nodes in $k$. 
The new integration weights $\Delta k$, and discretized kernel
are then rebuilt on the retained $k$ grid. This refines the sampling to be tailored for
kernel response and reduces redundant PPS degrees of freedom. This reduces correlations
between the PPS modes during reconstruction, leading to an invertible reconstruction covariance 
as shown in our previous work. All further analyses are carried out on the sparse grid.

\section{Validation tests with feature injections}
\label{sec:battery}

\begin{figure*}[!tbp]
\centering
\includegraphics[width=0.94\linewidth]{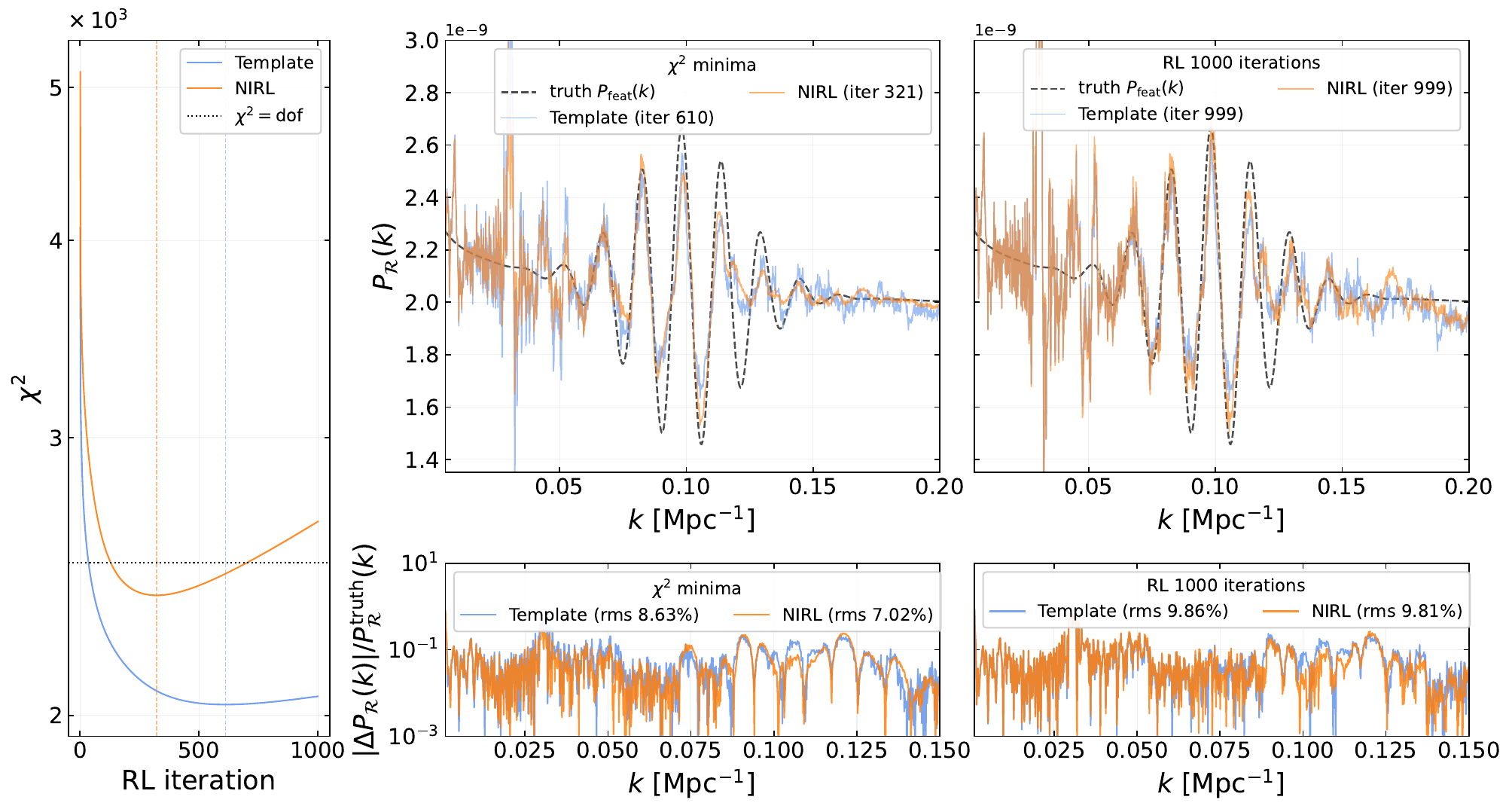}
\caption{Wavepacket reconstruction from one realisation using CV error estimates. Left:
Template and NIRL $\chi^2$ fit levels, with minima marked by dashed lines
and $N_\ell=2499$ dotted. Right: PPS reconstructions (upper panels) and absolute
fractional errors with respect to the truth PPS (lower panels), at the $\chi^2$ minima 
and at 1000 iterations, NIRL vs template. The same
bin size $9$ rolling mean is used for both methods as a regularization. Legend rms values are
over $0.001<k<0.2\,\mathrm{Mpc}^{-1}$.}
\label{fig:dfix_wp}
\end{figure*}

\begin{figure}[!tbp]
\centering
\includegraphics[width=0.90\linewidth]{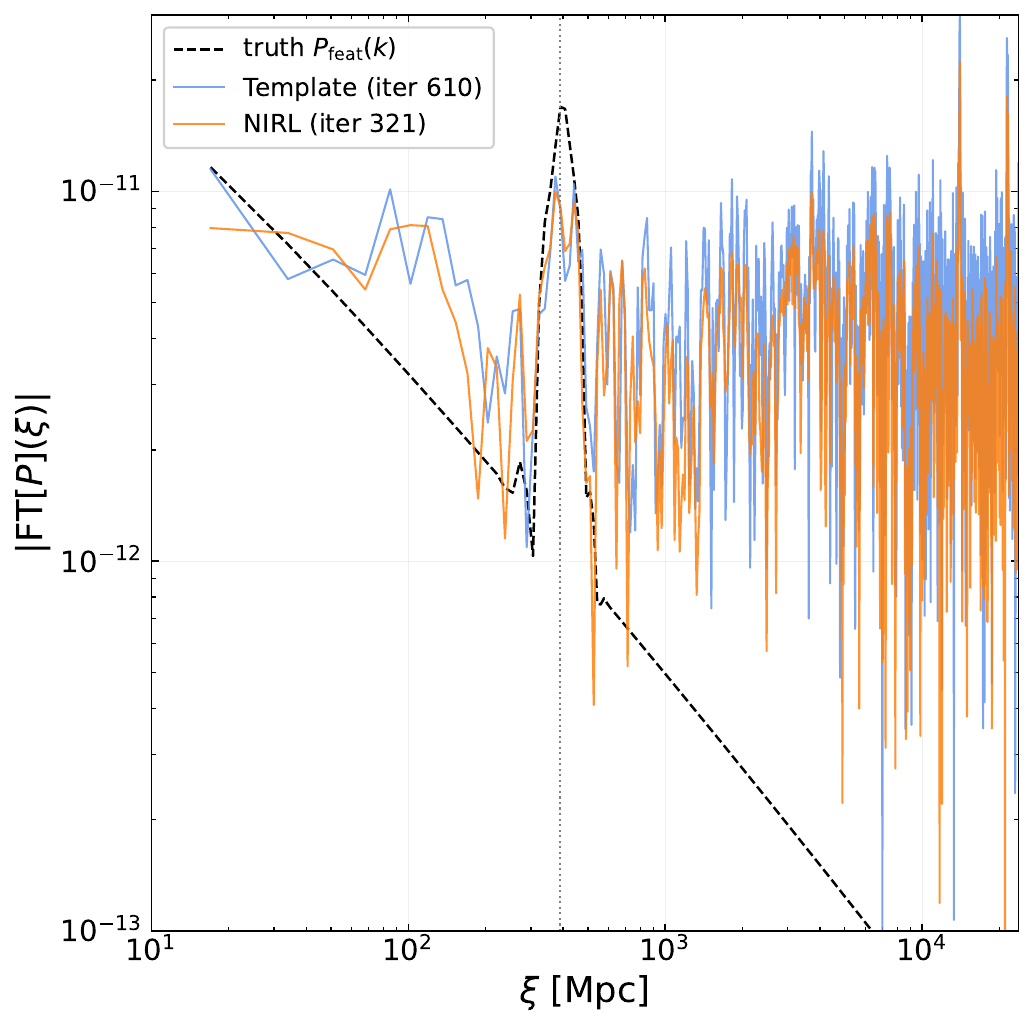}
\caption{Fourier spectra of the unsmoothed wavepacket for NIRL, template reconstructions
at the $\chi^2$ minima and for the injected truth PPS.
The feature peak near $\xi\simeq400$\,Mpc is visible above the noise floor, while the
reconstructions have excess broad-band "white noise" like power above approximately
$10^3$\,Mpc where kernel support is weak.}
\label{fig:dfix_ft_wp}
\end{figure}

\begin{figure*}[!tbp]
\centering
\includegraphics[width=0.94\linewidth]{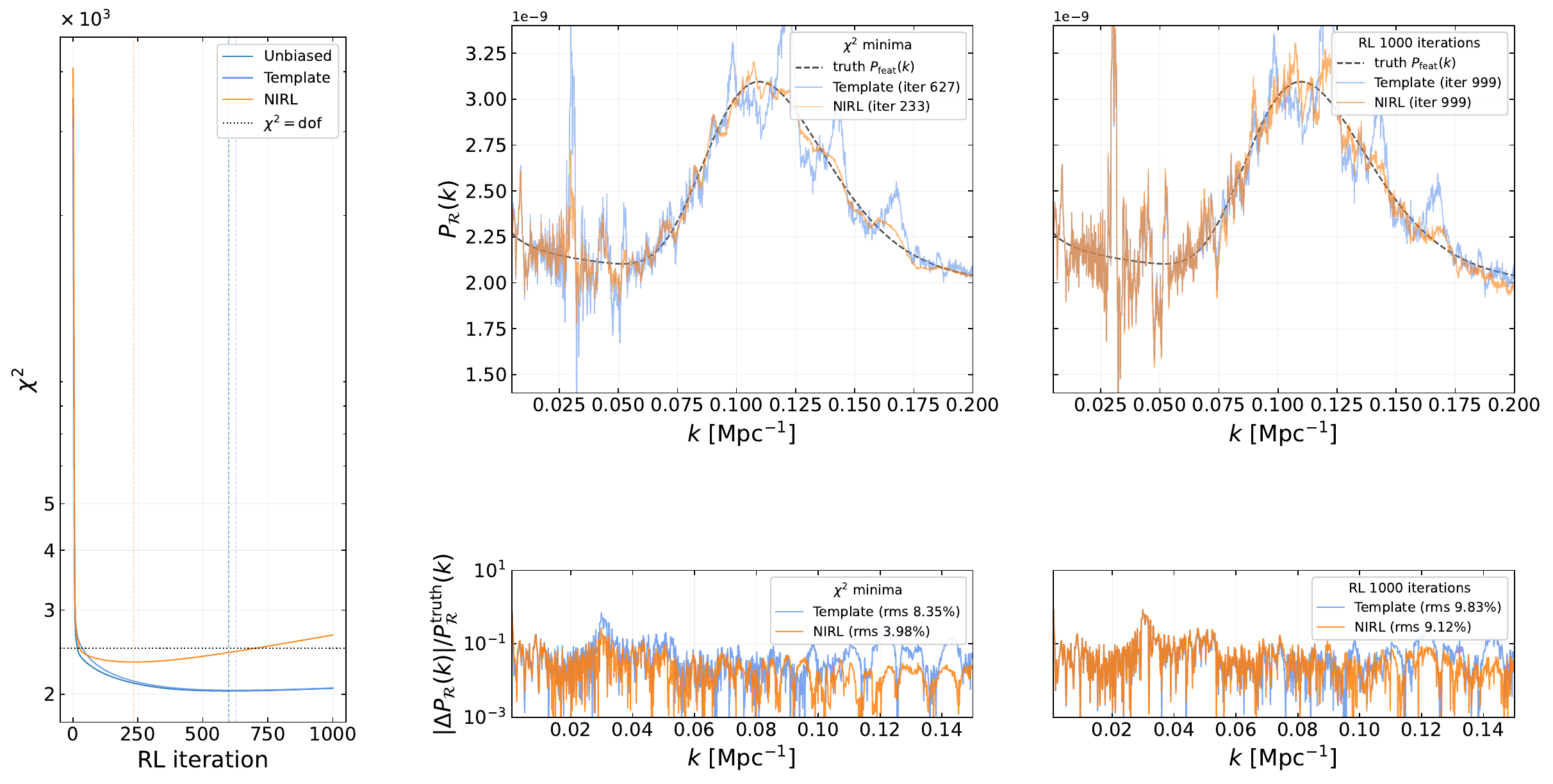}
\caption{Gaussian bump reconstruction from one realisation using CV error estimates. Left:
Template and NIRL $\chi^2$ fit levels, with minima marked by dashed lines
and $N_\ell=2499$ dotted. Right: PPS reconstructions (upper panels) and absolute
fractional errors with respect to the truth PPS (lower panels), at the $\chi^2$ minima 
and at 1000 iterations, NIRL vs template. The same
bin size $9$ rolling mean is used for both methods as a resgularization. Legend rms values are
over $0.001<k<0.2\,\mathrm{Mpc}^{-1}$. The lensing leakage in template reconstruction is
more strongly visible here.}
\label{fig:dfix_bump}
\end{figure*}

\begin{figure}[!tbp]
\centering
\includegraphics[width=0.90\linewidth]{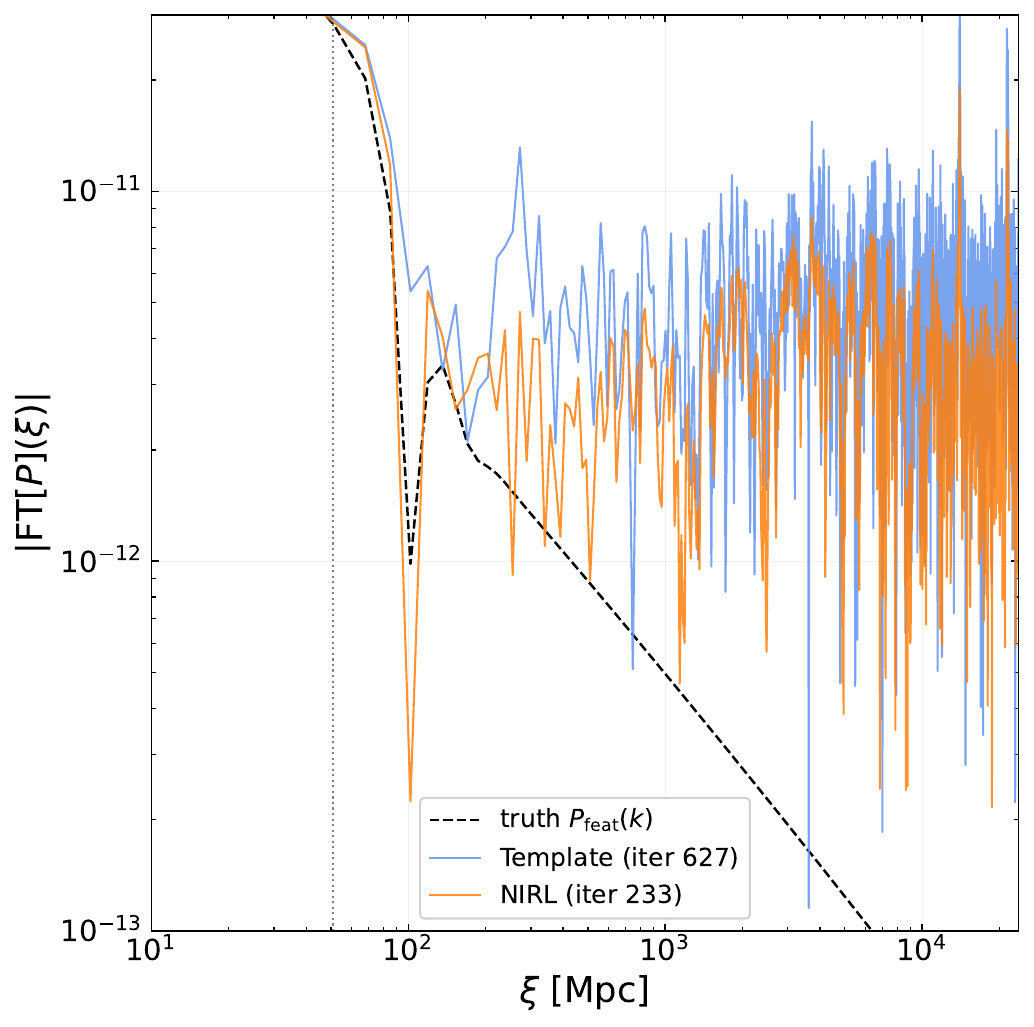}
\caption{Fourier spectra of the unsmoothed Gaussian bump for NIRL, template reconstructions
at the $\chi^2$ minima and for the injected truth PPS.
The smoother Gaussian spectral power is concentrated at lower Fourier frequencies, making the 
lensing leakage features more prominent at $\xi\simeq300$\,Mpc for template. The
reconstructions shows a similar broad "white noise" like excess at high $\xi$.
These modes motivate the regularization tested in
Sec.~\ref{sec:regularization}.}
\label{fig:dfix_ft_bump}
\end{figure}

We validate the NIRL estimator with a set of five input PPS features with distinct deviations from a power law form.
The features are mathematically designed to be within the $0.01<k<0.2\,\mathrm{Mpc}^{-1}$ range which is sensitive to lensing, 
and to probe a wide range 
of Fourier modes over $k$, which means we test for both slowly and rapidly oscillating features. In addition the power of the features
are designed to be such that their excess lensing contribution beyond the power law, lies within the cosmic variance envelope. 
This means that NIRL is validated for its ability to reconstruct beyond power law features that are roughly on 
the borderline limit of detection, making it a strong validation criterion. 
Five such features are shown in Fig.~\ref{fig:pkinput_all}, 
an oscillatory wavepacket, a soft wavepacket dominated by a dip and peak, an excess Gaussian bump, an upper wavepacket
containing excess power deviations, and a periodic non-sinusoidal feature with sharp edges.
These should not be considered as depictions of any particular inflationary models, but rather, analogs that test the algorithm numerically. 
Their fractional lensing contributions from beyond power law features are shown in Figure~\ref{fig:features_overlay}. 
The Gaussian bump produces the largest overall change while oscillatory features are less pronounced at the $C_\ell$ level, since they
can partly cancel out power when convolved with the lensing kernel.

We will use the following terminology to distinguish the different lensing reconstruction handling.
\emph{Unbiased} is the IRL reconstruction from the corresponding and ideal unlensed spectrum as a control group.
\emph{Biased} is the IRL applied to lensed power without any lensing correction (template or NIRL). 
This is also an "antagonistic" control group to show the effect of lensing as a systematic, on the reconstructed PPS. This is 
not directly tested or shown here, since it is already well known in the literature that lensing produces spurious 
oscillatory features in the reconstructed PPS if unaccounted. We will focus on comparing NIRL to template with the understanding
that if NIRL performs better than template, it automatically corrects for the \emph{Biased} scenario as well.
\emph{Template} is the IRL reconstruction after subtraction of a fixed power law prior sourced lensing correction, from the lensed data, 
which is the conventional approach in existing literature.
\emph{NIRL} is our reconstruction that updates the lensing kernel iteratively in situ from the previous iteration PPS.
The reconstruction is carried out on noiseless data, as a validation for the estimator, hence the $tanh$ error weighting 
in Eq.~\eqref{eq:IRL_eqns} is not applied, nor is there a $\chi^2$ convergence term in this section. Hence we perform the IRL and NIRL 
reconstructions for a fixed number of iterations at 500, which is sufficient for convergence of the estimator on noiseless data.

Figures~\ref{fig:pkrecon_all} and \ref{fig:pkreconerr_all} compare the
reconstructions at 500 iterations for the features. 
To provide a single statistic evaluation of NIRL delensing, we quote
the rms percentage difference with respect to the unbiased reconstruction for both template IRL and NIRL, using the formula

\begin{equation}
 \text{rms} =100\left[\frac{1}{N_{k}}
 \sum_{k_i}
 \left(\frac{P_X(k_i)-P_{\rm U}(k_i)}{P_{\rm U}(k_i)}\right)^2
 \right]^{1/2},
 \label{eq:feature_rms}
\end{equation}
where $X$ is reconstruction type and $P_{\rm U}$ is the unbiased benchmark.

\begin{table}[htb]
\centering
\begin{tabular}{lrrr}
\toprule
Feature & Iteration & Template (\%) & NIRL (\%) \\
\midrule
Wavepacket                & 500 & 2.09 & 1.18 \\
Soft wavepacket           & 500 & 0.97 & 0.33 \\
Gaussian bump             & 500 & 3.24 & 0.59 \\
Upper wavepacket          & 500 & 1.10 & 0.21 \\
Periodic non-sinusoidal    & 500 & 0.99 & 0.46 \\
\bottomrule
\end{tabular}
\caption{Rms percentage errors with respect to the unbiased reconstruction for
the five features depicted in Fig.~\ref{fig:pkrecon_all} using
Eq.~\eqref{eq:feature_rms}.}
\label{table:battery}
\end{table}

Overall it is evident that NIRL reduces the lensing systematic in the reconstruction
for every type of feature. The individual per feature rms values are tabulated in Table~\ref{table:battery}.
We observe that the reconstruction accuracy post NIRL lensing debiasing is more effective for features that are less oscillatory in nature, 
implying that lensed deconvolution kernel has overall fundamental sensitivity limits to different Fourier modes over $k$.
This motivates our next analysis proposal, which is to validate the estimator in the presence of noise, implement a robust regularization technique,
and quantify its reconstruction error.

\section{Reconstruction validation under observation errors}
\label{sec:noise}

\begin{figure}[!tbp]
\centering
\begin{subfigure}{\linewidth}
  \includegraphics[width=\linewidth]{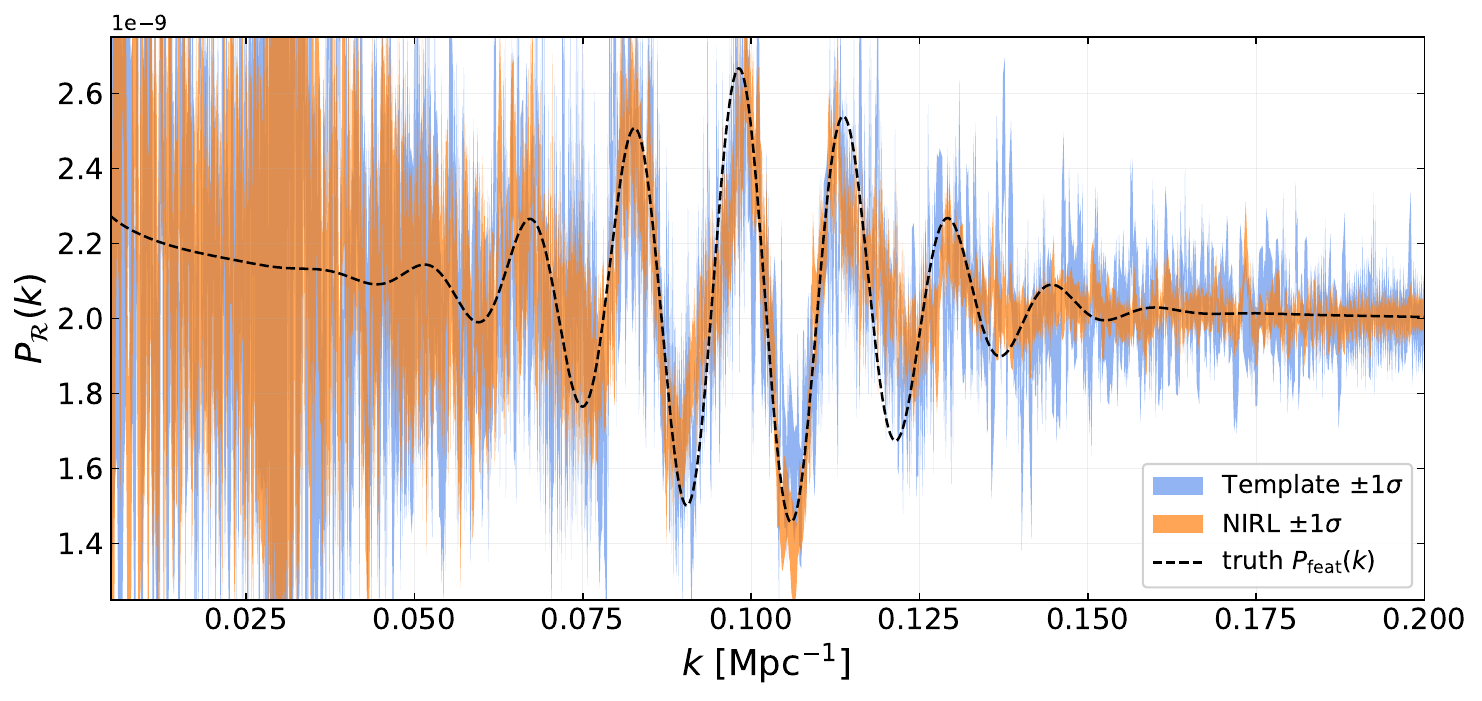}
  \caption{}
  \label{fig:boot_wp_1bin_a}
\end{subfigure}\par\smallskip
\begin{subfigure}{\linewidth}
  \includegraphics[width=\linewidth]{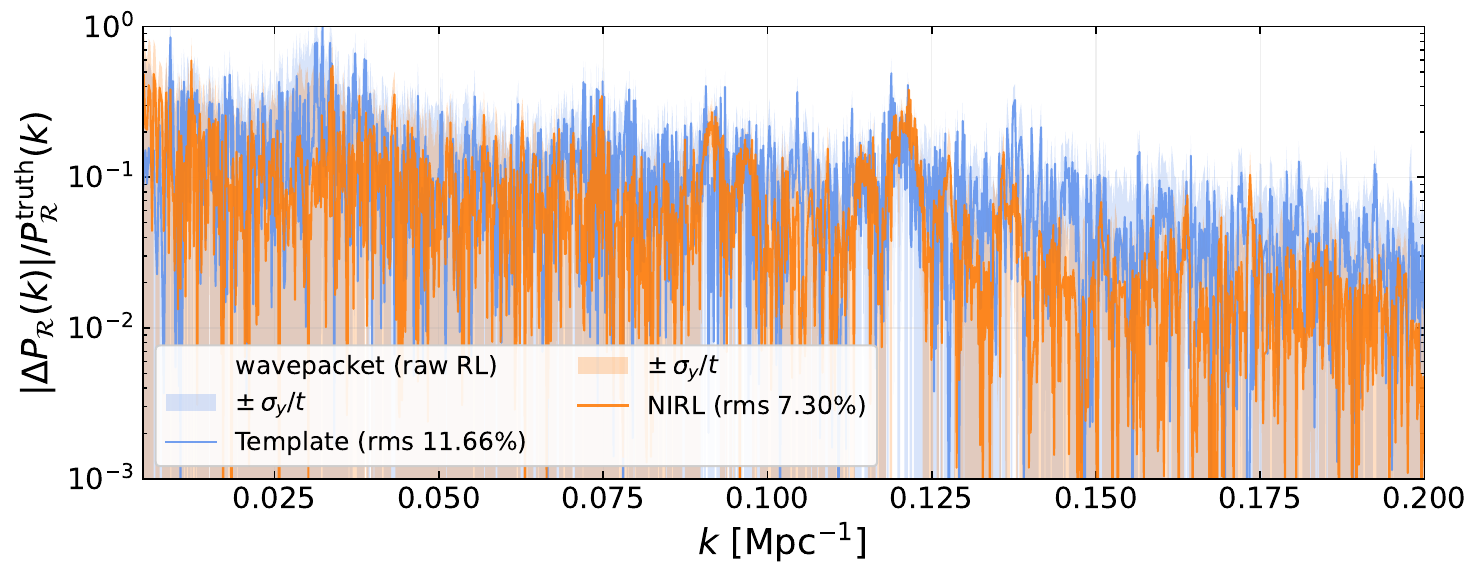}
  \caption{}
  \label{fig:boot_wp_1bin_b}
\end{subfigure}
\caption{Template vs NIRL wavepacket bootstrap with no binning.
(a) $1\sigma$ bands centred on one representative realisation.
(b) Absolute fractional errors and corresponding $1\sigma$ bands. 
Legend rms values use $0.02<k<0.3\,\mathrm{Mpc}^{-1}$.}
\label{fig:boot_wp_1bin}
\end{figure}

\begin{figure}[!tbp]
\centering
\begin{subfigure}{\linewidth}
  \includegraphics[width=\linewidth]{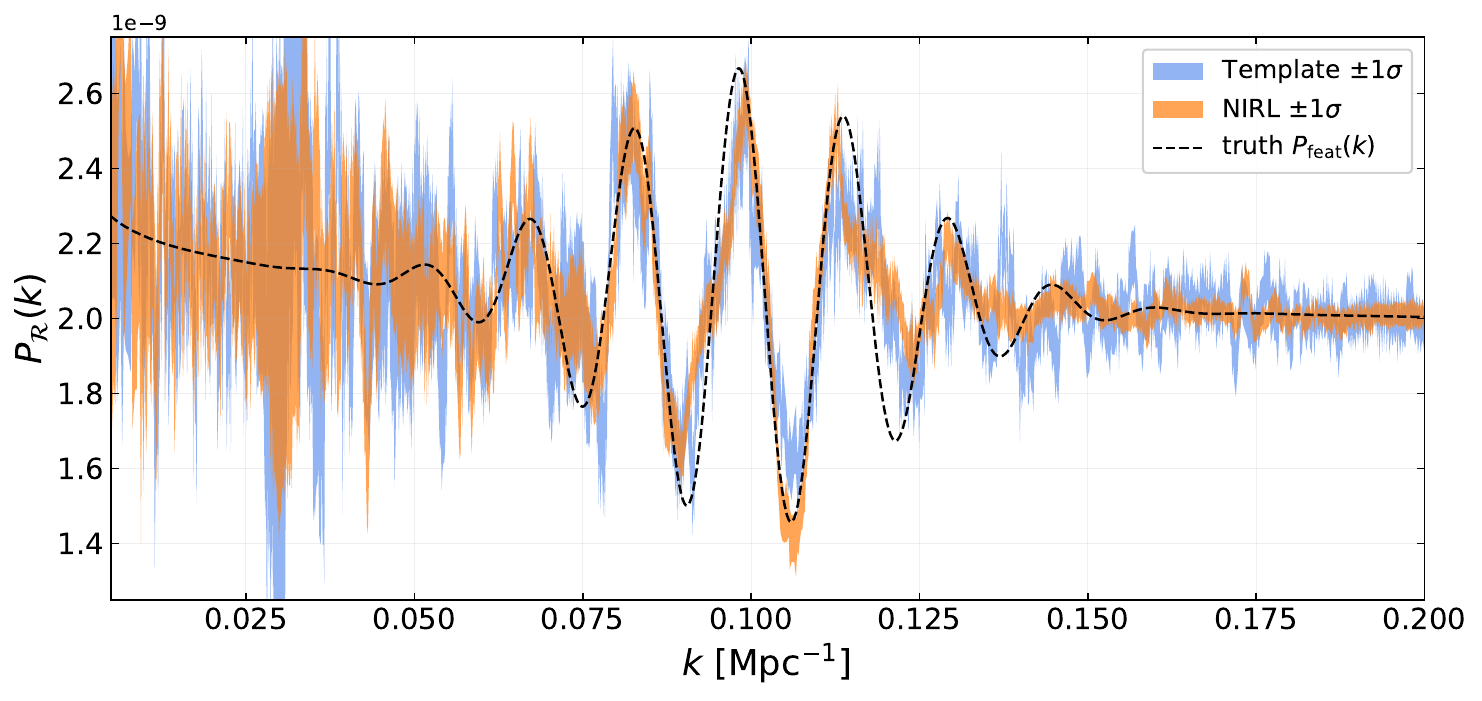}
  \caption{}
  \label{fig:boot_wp_3bin_a}
\end{subfigure}\par\smallskip
\begin{subfigure}{\linewidth}
  \includegraphics[width=\linewidth]{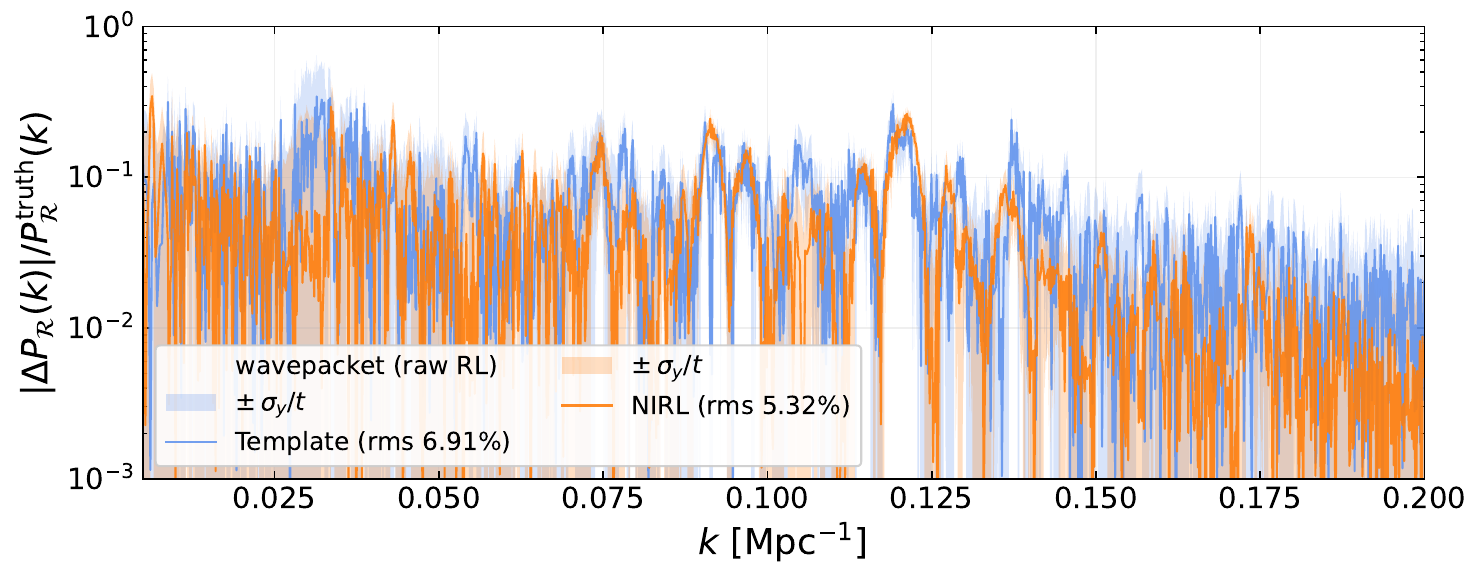}
  \caption{}
  \label{fig:boot_wp_3bin_b}
\end{subfigure}
\caption{Template vs NIRL wavepacket bootstrap with a bin size $3$ rolling mean.
(a) $1\sigma$ bands centred on one representative realisation.
(b) Absolute fractional errors and corresponding $1\sigma$ bands. 
Legend rms values use $0.02<k<0.3\,\mathrm{Mpc}^{-1}$. The larger averaging window reduces scatter
but may also smooth reconstructed features arbitrarily.}
\label{fig:boot_wp_3bin}
\end{figure}

\begin{figure}[!tbp]
\centering
\begin{subfigure}{\linewidth}
  \includegraphics[width=\linewidth]{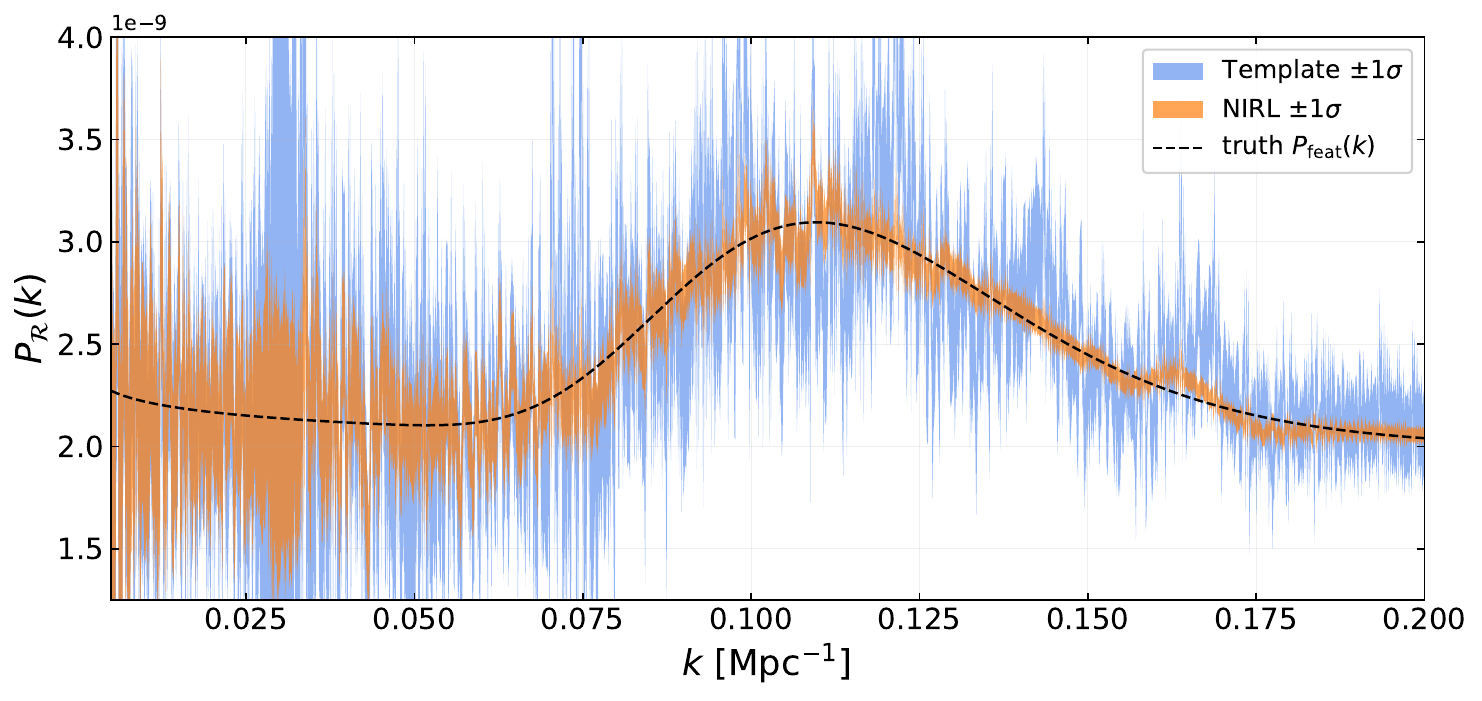}
  \caption{}
  \label{fig:boot_bump_1bin_a}
\end{subfigure}\par\smallskip
\begin{subfigure}{\linewidth}
  \includegraphics[width=\linewidth]{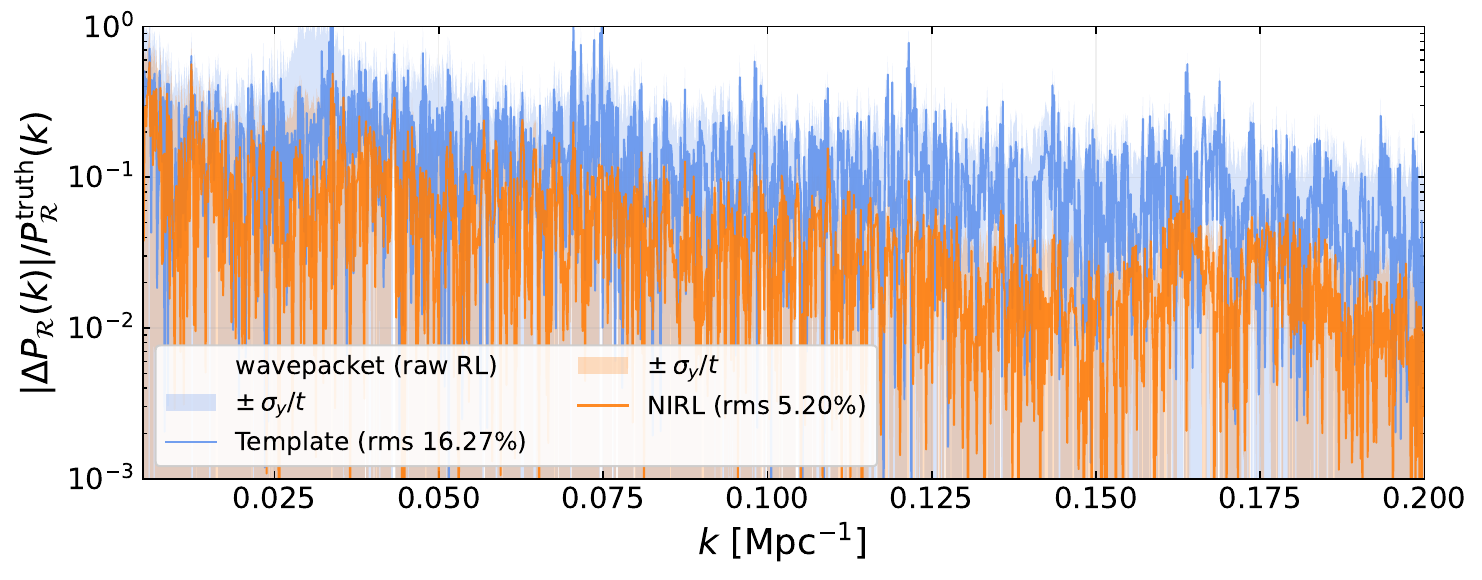}
  \caption{}
  \label{fig:boot_bump_1bin_b}
\end{subfigure}
\caption{Template vs NIRL Gaussian bump bootstrap with no binning.
(a) $1\sigma$ bands centred on one representative realisation.
(b) Absolute fractional errors and corresponding $1\sigma$ bands. 
Legend rms values use $0.02<k<0.3\,\mathrm{Mpc}^{-1}$.}
\label{fig:boot_bump_1bin}
\end{figure}

\begin{figure}[!tbp]
\centering
\begin{subfigure}{\linewidth}
  \includegraphics[width=\linewidth]{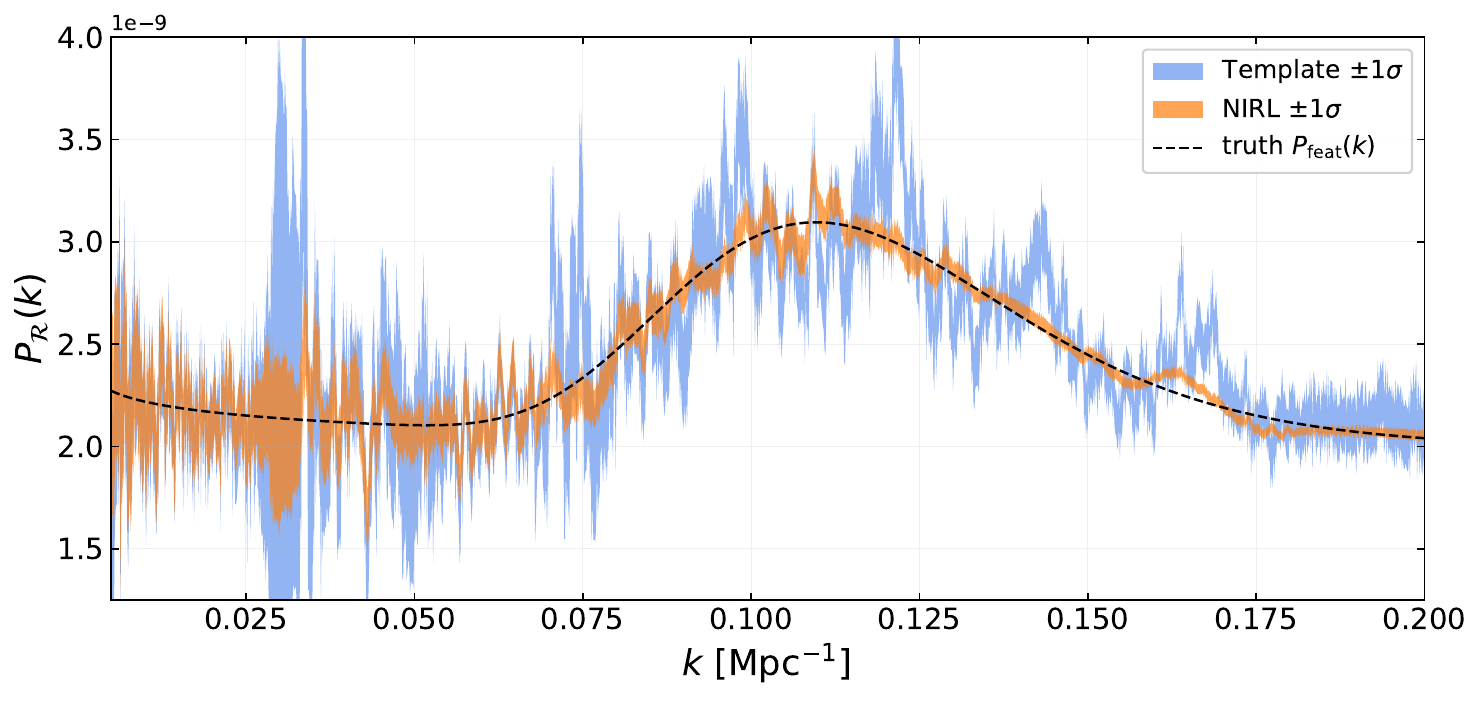}
  \caption{}
  \label{fig:boot_bump_3bin_a}
\end{subfigure}\par\smallskip
\begin{subfigure}{\linewidth}
  \includegraphics[width=\linewidth]{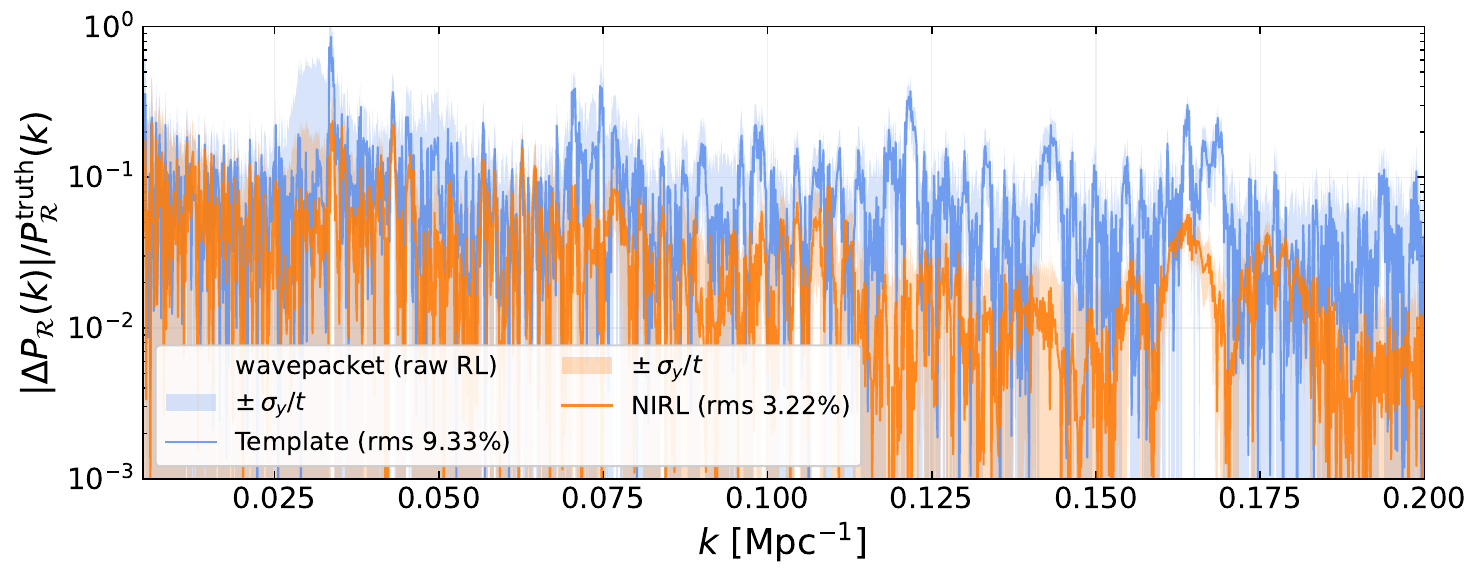}
  \caption{}
  \label{fig:boot_bump_3bin_b}
\end{subfigure}
\caption{Template vs NIRL Gaussian bump bootstrap with a bin size $3$ rolling mean.
(a) $1\sigma$ bands centred on one representative realisation.
(b) Absolute fractional errors and corresponding $1\sigma$ bands. 
Legend rms values use $0.02<k<0.3\,\mathrm{Mpc}^{-1}$. 
Consistently reduced variance and unbiasedness is observed in NIRL again, 
due to the feature being more isolated from noisy modes in Fourier space.}
\label{fig:boot_bump_3bin}
\end{figure}

\subsection{Cosmic-variance simulations and convergence prescription}

We use cosmic variance limited error bars for the feature $C_{\ell,\rm truth}^{TT}$ given by
\begin{equation}
 \sigma_\ell=\sqrt{\frac{2}{2\ell+1}}\,C_{\ell,\rm truth}^{TT},
 \qquad 2\leq\ell\leq2500.
 \label{eq:cv_sigma}
\end{equation}
and generate independent Gaussian realisations where the 
$C_{\ell,\rm truth}^{TT}$ is the appropriate lensed input
spectrum as the mean and $\sigma^2_\ell$ for the variance. 
Template and NIRL use the same lensed data realisation, and we restore
the MRL $tanh^2$ weighting of Eq.~\eqref{eq:IRL_eqns}. 
The Gaussian draw approximates the non-Gaussian distribution of low-$\ell$ bandpowers, 
however this is not an issue for our tests which are sensitive to high $k$ and correspondingly high $\ell$ regimes.

Unlike before, we now perform the template vs NIRL reconstruction for upto 1000 iterations
to allow a maximal fit of any features hidden in the data within the noise, 
while also plotting the $\chi^2$ statistic to select the best fit PPS.

\subsection{Iteration dependence under noisy reconstruction}
\label{sub:iterfix}

 The figures~\ref{fig:dfix_wp} and \ref{fig:dfix_bump} show the $\chi^2$ reconstruction fit 
 to the input data $C_{\ell,\rm rlzn}^{TT}$ realisation for the wavepacket and Gaussian bump features,
 comparing template and NIRL results respectively, for upto 1000 iterations. 
 They also plot the absolute fractional rms error with respect to the truth PPS at the selected $\chi^2$ minima iteration and at 1000 iterations.
 The PPS curves are binned with a rolling mean of bin size 9.

It can be seen that at the $\chi^2$ minima iteration, the absolute rms error is $7.02\%$ for NIRL
and $8.63\%$ for Template for the wavepacket, and $3.98\%$ versus $8.35\%$
for the Gaussian bump. This harkens to our earlier experience that while NIRL consistently 
debiases better, the fidelity is also sensitive to the Fourier modes being probed. We note that the NIRL 
reaches the $\chi^2$ minima faster than template, while being less biased than template. This suggests that 
the template struggles to fit the data and ends up overfitting noise as well as the lensing bias, mixing the two effects.
We conclude that NIRL is more strongly unbiased and also manages noise better.

We also note that by iteration 1000 the rms has increased to $9.81\%$ and $9.86\%$ for
the wavepacket, and $9.12\%$ and $9.83\%$ for the bump, respectively.
While for the wavepacket case, we see that at 1000 iterations the visual reconstruction of the features
are more accurate than at the $\chi^2$ minima, the rms error as well as noise overfit are higher for NIRL.
In contrast, the Gaussian bump case shows that the $\chi^2$ minima coincides with a visually more accurate 
fit to the truth PPS using NIRL, while at 1000 iterations, NIRL starts to overfit noise at the expense of the feature reconstruction.
This is a key observation; even as NIRL consistently remains less biased in terms of rms error than template, the 
stopping criteria now becomes constrained by noise overfitting and the sensitivity of the kernel to the deconvolution at different Fourier modes.
It now motivates the regularization algorithm, where we will introduce a Fourier domain filtering 
to mitigate noise overfitting at high $\xi$ modes where kernel support is weak,
while retaining the low $\xi$ modes that are sensitive to the features of our interest.

\subsection{Bootstrap precision and the role of binning}
\label{sub:bootstrap}

\begin{figure*}[!tbp]
\centering
\includegraphics[width=0.94\linewidth]{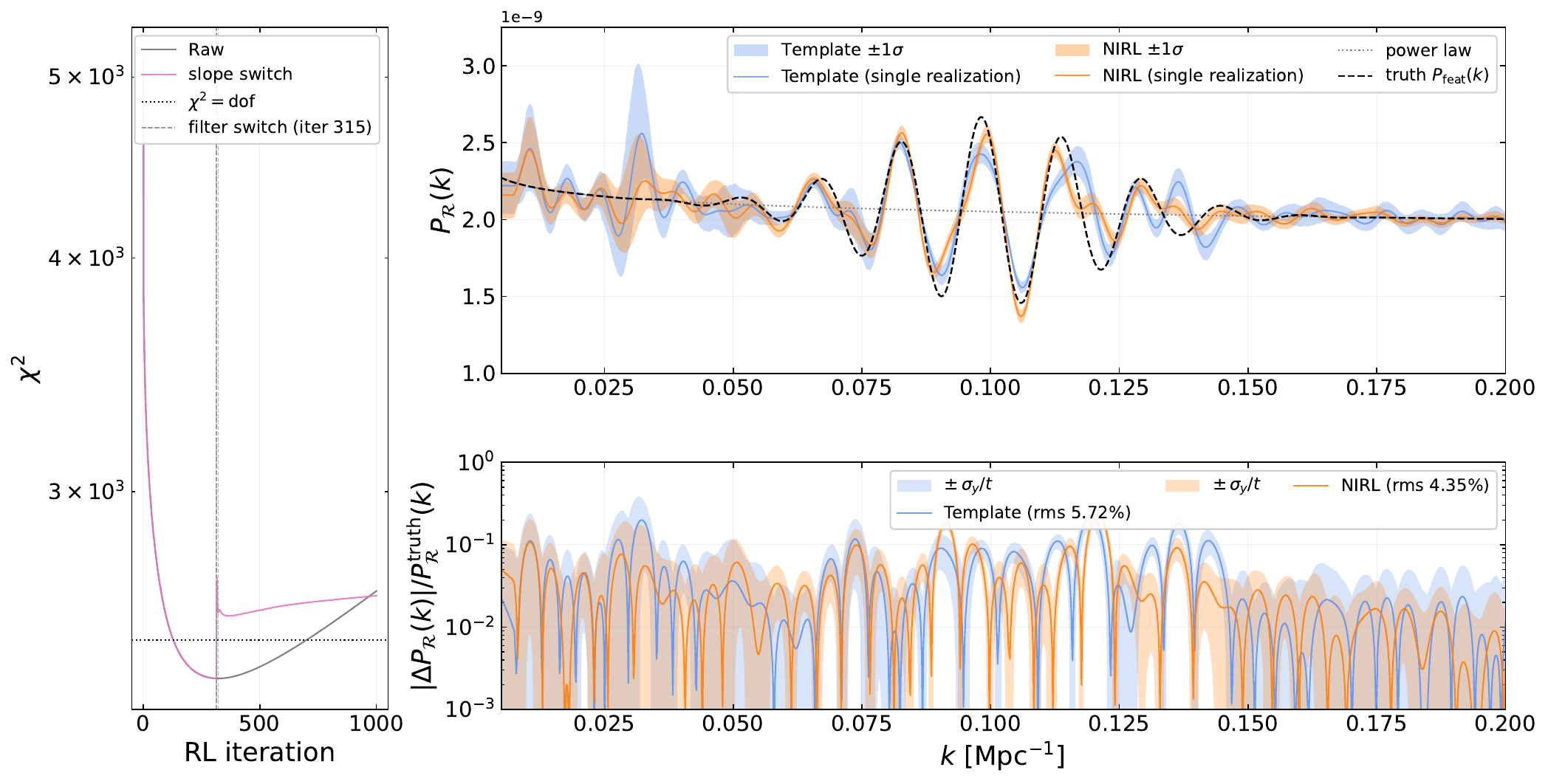}
\caption{FLIP wavepacket test. Left: Unfiltered vs filtered NIRL
$\chi^2$ levels over 1000 iterations, with filter activation at
minima iteration 315. After switch, the filtered curve ascends slower than the unfiltered
and crosses below it near iteration 969. The dotted level is
$N_\ell=2499$. Right, upper: filtered Template and NIRL PPS reconstruction
for a representative realisation, with forward modelled $\sigma$ bands, and the true
PPS. Right, lower: absolute fractional errors with the corresponding
$\sigma$ bands. Legend rms values use
$0.02<k<0.3\,\mathrm{Mpc}^{-1}$.}
\label{fig:fftboot_wp}
\end{figure*}

\begin{figure}[!tbp]
\centering
\includegraphics[width=0.90\linewidth]{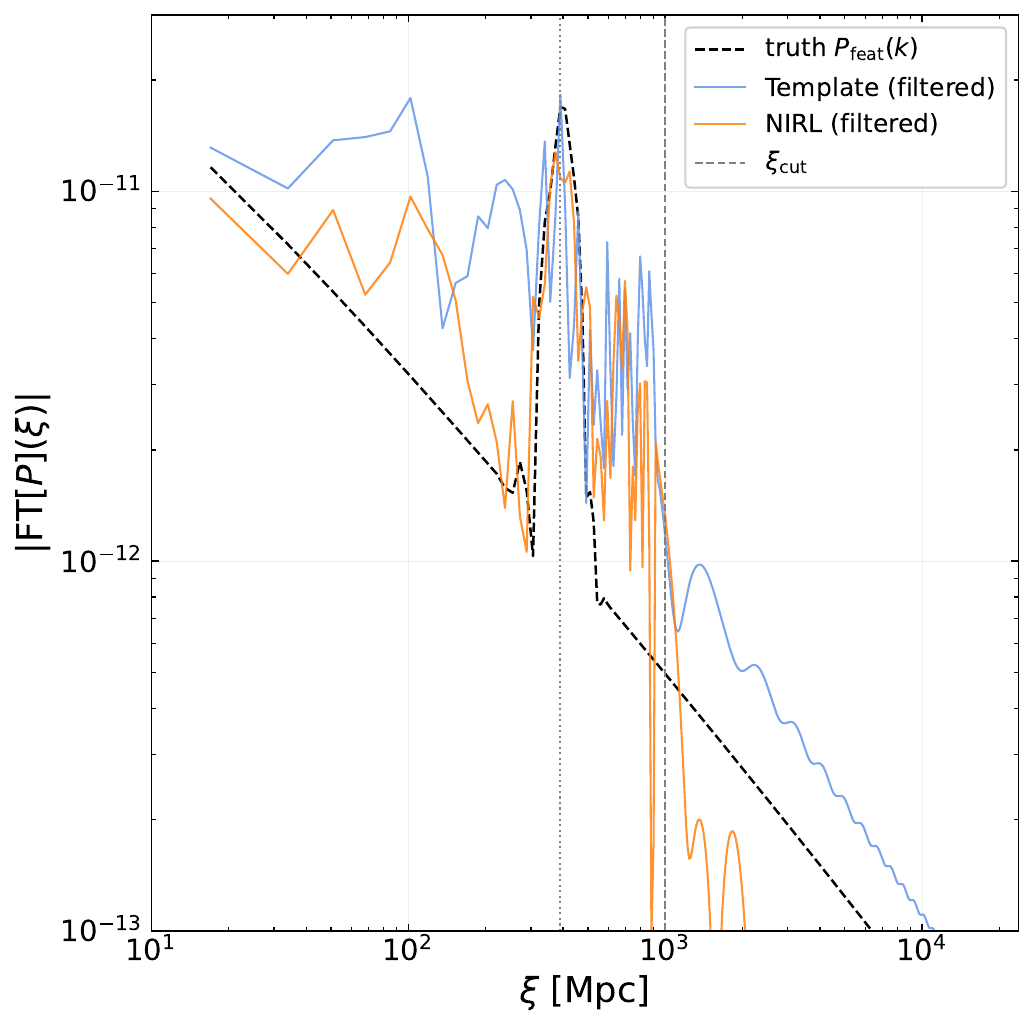}
\caption{Fourier spectra of the FLIP filtered wavepacket reconstructions.
The dashed line marks $\xi_c=10^3$\,Mpc. The feature peak near
$400$\,Mpc remains, while the high mode broadband noise excess is strongly suppressed.
Note that the low mode excess is also reduced for NIRL, showing how the NIRL unbiasing and FLIP 
are complementary with respect to noise control.}
\label{fig:fftboot_ft_wp}
\end{figure}

\begin{figure*}[!tbp]
\centering
\includegraphics[width=0.94\linewidth]{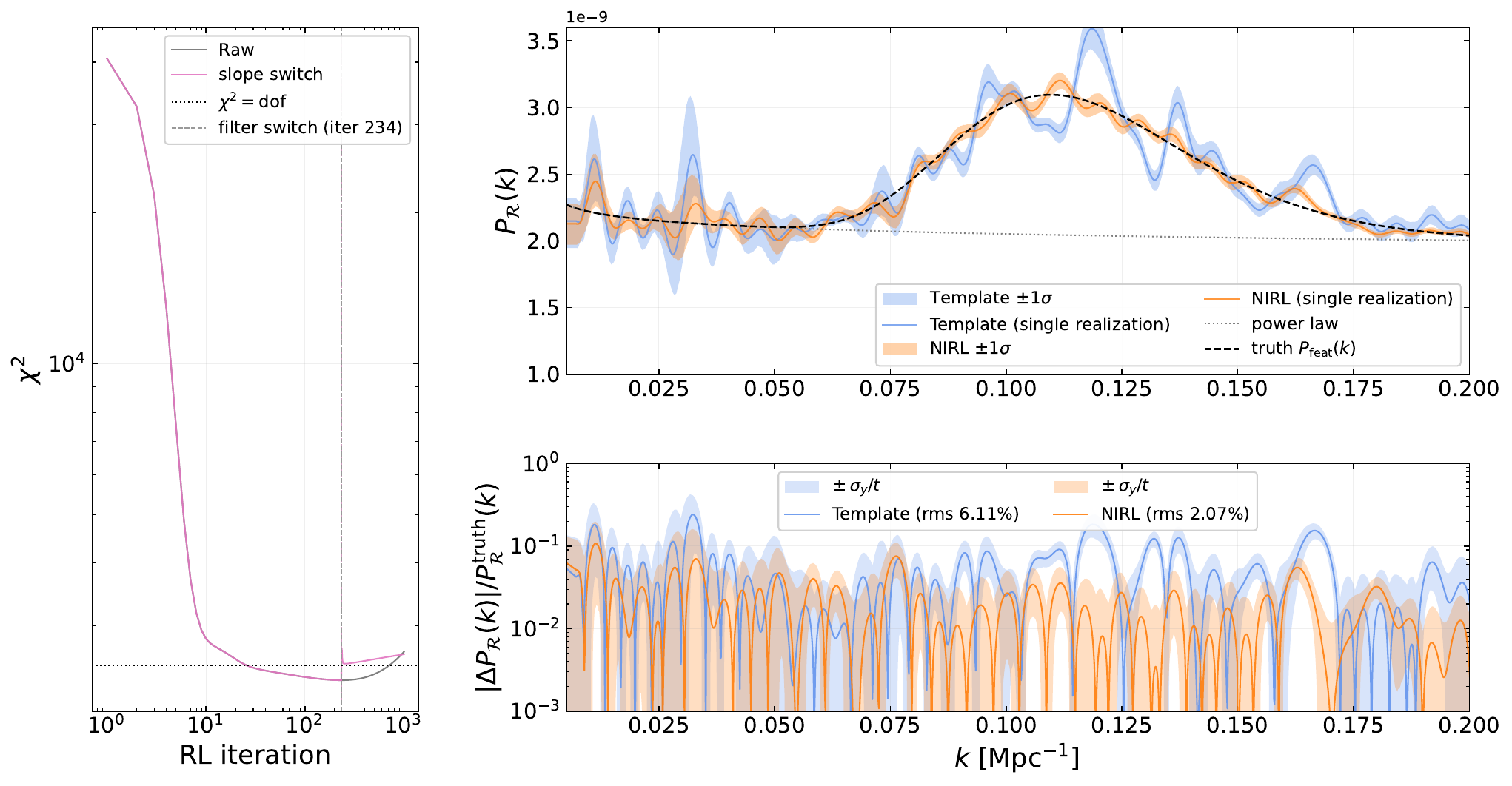}
\caption{FLIP Gaussian bump test. Left: Unfiltered vs filtered NIRL
$\chi^2$ levels over 1000 iterations, with filter activation at
minima iteration 234. After switch, the filtered curve ascends slower than the unfiltered
and crosses below it near iteration 941. The dotted level is
$N_\ell=2499$. Right, upper: filtered Template and NIRL for a representative realisation
PPS estimate, with forward modelled $\sigma$ bands, and the true
PPS. Right, lower: absolute fractional errors with the corresponding
$\sigma$ bands. Legend rms values use
$0.02<k<0.3\,\mathrm{Mpc}^{-1}$.}
\label{fig:fftboot_bump}
\end{figure*}

\begin{figure}[!tbp]
\centering
\includegraphics[width=0.90\linewidth]{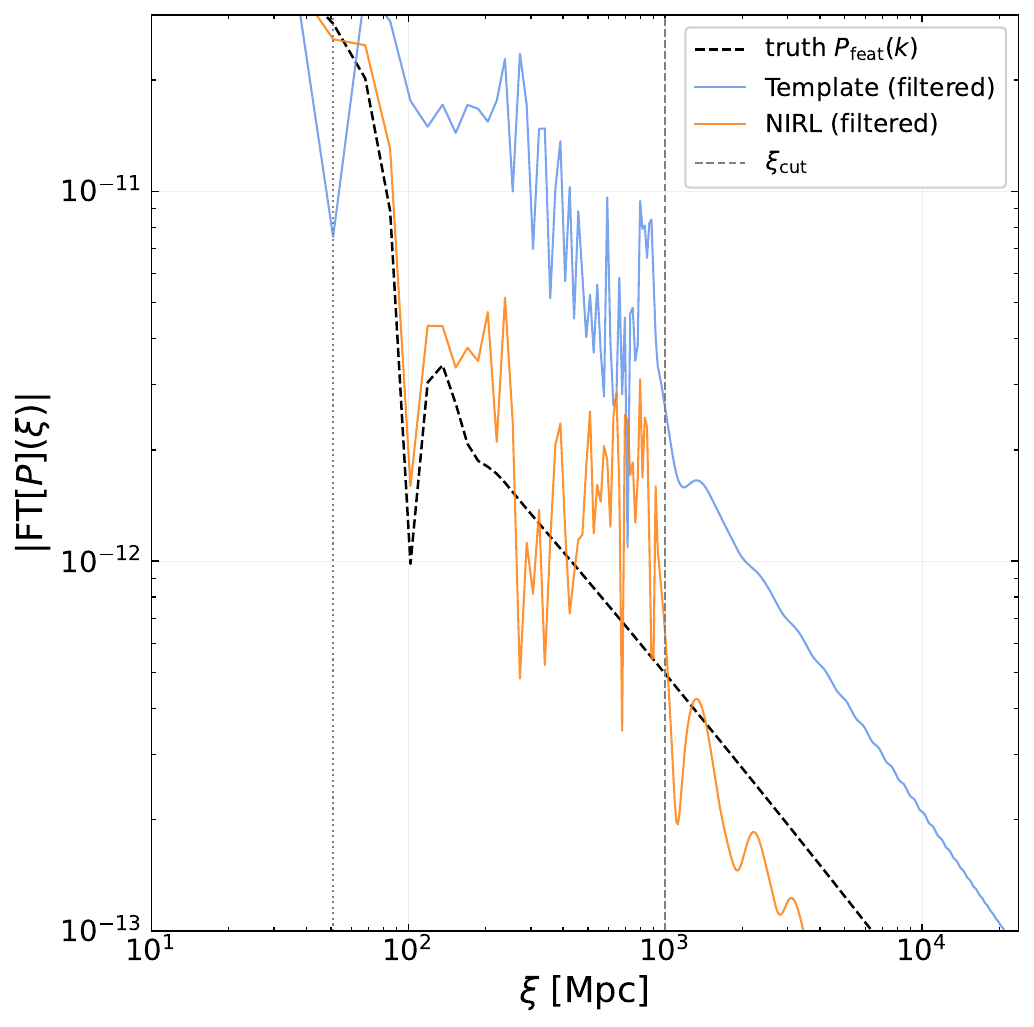}
\caption{Fourier spectra of the FLIP filtered Gaussian bump reconstructions.
The dashed line marks $\xi_c=10^3$\,Mpc. The feature trough near
$200$\,Mpc remains, while the high mode broadband excess is significantly suppressed.
Significant improvement in the reconstruction fidelity is observed, as well as suppression
of power excess at both low and high $\xi$ modes.}
\label{fig:fftboot_ft_bump}
\end{figure}

In this section we quantify the reconstruction precision for the wavepacket and Gaussian-bump features, 
using the same CV realisations as in Sec.~\ref{sub:iterfix}. We draw 1000 realizations of the lensed $C_\ell^{TT}$ for each feature,
perform the template versus NIRL reconstruction as before, and evaluate the ensemble variance of the reconstructed PPS, and plot
the standard deviation around a single representative realisation. This time we display the plots in Figures~\ref{fig:boot_wp_1bin} and \ref{fig:boot_wp_3bin} 
for the wavepacket, and Figures~\ref{fig:boot_bump_1bin} and \ref{fig:boot_bump_3bin} for the Gaussian bump, to show how 
the rolling mean binning affects the reconstruction scatter.
As can be seen while the reconstruction unbiasing remains consistent, the scatter is reduced with larger binning, 
but at the cost of smoothing out the features of interest. Without a consistent or physically motivated binning or regularization
prescription it becomes difficult to quantify the reconstruction precision. This is a short section to set up the reconstruction precision 
pipeline, and we elaborate with a novel regularization technique in the next section.

\section{Regularization}
\label{sec:regularization}

\subsection{Motivation for Fourier filtering}
\label{sub:filter_motivation}

Summarizing the results of the previous sections, we have seen that the NIRL estimator is a robust and consistent
debiasing estimator for the lensing systematic. Once we add noise to the reconstruction, we run into two key 
concerns; stopping criteria for the iterations, and more importantly, a regularization scheme to reduce reconstruction
noise while retaining reconstruction fidelity. To motivate our regularization prescription, we examine the Fourier 
spectra of the noisy single realization reconstructions for the wavepacket and Gaussian bump features, depicted unsmoothed 
in Figs.~\ref{fig:dfix_ft_wp} and \ref{fig:dfix_ft_bump}.

Throughout this section the transform is with respect to $k$,
\begin{equation}
 \widehat P(\xi)=\int P(k)e^{-i\xi k}\,dk,
 \label{eq:pk_fourier}
\end{equation}
so $\xi$ has units of Mpc.

We can clearly see that for both features, we can see a possible flat spectrum "white noise"
like region beyond the injected features Fourier spectra in the dashed black lines, above approximately $10^3$\,Mpc. 
The injected wavepacket has a distinct peak near
$\xi\simeq400$\,Mpc, and the bump is also concentrated at lower $\xi$ modes.
We note that despite the noise overfitting at higher $\xi$ modes, the NIRL estimator 
continues to be nearly an order of magnitude less noisy as well as unbiased than the template,
clearly visible in the visual separation of the orange NIRL and blue template spectral power. 

If we observe Figure~\ref{fig:ft_glk} it presents the convolution kernel in the
Fourier domain as well, and we can see that the kernel is sensitive to low $\xi$ modes, 
but tends to drop in sensitivity beyond $\xi\simeq10^3$\,Mpc, for most $\ell$ bands, 
and especially so for the higher $\ell$ bands that are most sensitive to lensing, as well as higher $k$ modes
that we are probing for power law departures, based on both predicted inflation models as well as lower cosmic variance.

We conclude that due to the inherently non linear nature of the RL estimator, 
the estimator "leaks" noise from any given $\ell$ data point, across all $k$ modes depending on the 
Fourier sensitivity of the kernel. This is a fundamental limitation of the RL estimator, which has been
demonstrated in Liu et al.\ \cite{RLnature2025}. The
Cram\'er--Rao lower bound (CRLB) on the variance becomes unbounded where the transfer response is small.
Based on this we propose a simple but effective regularization scheme,
where we perform a low pass filter of the reconstructed PPS in the Fourier domain at each iteration
step to suppress the noise leakage at higher Fourier modes that are unsupported by an appropriate kernel sensitivity.
The cutoff can now be chosen on a physically motivated basis, and we choose $\xi_c=10^3$\,Mpc, 
which is the approximate scale where the kernel sensitivity drops by roughly an order of magnitude 
for most of the higher $\ell$ bands relative to our cosmology. We should note that
this does not evade the runaway CRLB limitation but is a physically motivated regularization scheme to mitigate noise overfitting, not solve it.
Other regularization schemes also perform similar roles, by penalizing either the roughness of the reconstruction
or minimizing total variation, but we note that our scheme optimizes more directly for the net available information content in Fourier domain. 
To visualize what this means, in Figs.~\ref{fig:dfix_ft_wp} and \ref{fig:dfix_ft_bump}, 
any features in the Fourier domain that are indistinguishable from a flat "noise" like spectrum, cannot be recovered,
at least not without some prior assumptions on either the noise spectrum or expected features. Our regularization
is a frequentist approach where we aim to suppress such noisy modes based on the physics of our convolution 
kernel and expectation of the underlying signal having a detectable SNR at the relevant scales. 
This way we avoid losing a significant amount of information the kernel is sensitive to, 
while suppressing the noise at other "nuisance" $\xi$ modes. 
We further note that most mainstream inflation models predict features that are not in Fourier modes
that are significantly higher 
than the kernel sensitivity region, and if they were, they would be in principle difficult to detect 
from any 2-D projection of the 3-D $P(k)$.

\subsection{Filtered iterations, bootstrap tests, and reconstruction results}
\label{sub:filter_tests}

We repeat the NIRL vs template reconstruction for the cosmic variance error based realization data 
for both the wavepacket and bump test features for a 1000 realizations each, but this time with the iterative 
Fourier Low Iterative Pass (FLIP) filter.
We start the reconstruction first without the FLIP filter and identify the first $\chi^2$ minimum.
Then we switch on the iterative FLIP filter and continue the reconstruction for higher iterations until
we reach a second $\chi^2$ minimum.
The reason for delayed first minimum activation is to allow the unfiltered reconstruction
to develop any existing features and reduce lensing bias, alongwith noise, before damping the noise leakage 
and allowing NIRL-FLIP to converge to a maximally unbiased and noise suppressed reconstruction.

For the actual filtering process, the PPS at each iteration after the first minimum, 
is interpolated onto a uniform $k$ grid (since the kernel sampling is in log space), 
and Fourier transformed using a Hann window.  

The FLIP filter can be depicted by 
$\widehat P_{\rm f}(\xi)=H(\xi)\widehat P(\xi)$ on the windowed spectrum,
using Fourier cutoff mode $\xi_c=10^3$\,Mpc and a cosine transition of width
$\delta\xi=0.08\xi_c$ where

\begin{equation}
 H(\xi)=
 \begin{cases}
 1, & |\xi|<\xi_c-\delta\xi,\\
 \frac12\!\left[1+\cos\!\left(\pi
 \frac{|\xi|-\xi_c+\delta\xi}{\delta\xi}\right)\right],
 & \xi_c-\delta\xi\leq|\xi|\leq\xi_c,\\
 0, & |\xi|>\xi_c.
 \end{cases}
 \label{eq:fourier_taper}
\end{equation}
After the inverse Fourier transform back to $k$ space, the filtered PPS is interpolated back to the sparse grid
and the next iteration update is carried out for both the lensing correction and then PPS in the case of NIRL
and only the PPS in the case of template. 

This procedure is performed for the 1000 CV realizations and the consequent $\chi^2$ histories
, FLIP filtered PPS reconstructions and the absolute rms errors with respect to the truth PPS 
are shown in Figs.~\ref{fig:fftboot_wp} and \ref{fig:fftboot_bump}. 
We show the $1\sigma$ and $2\sigma$ confidence intervals between NIRL and template for the FLIP filter.

We can see again that the NIRL-FLIP reconstruction is consistently less biased than template-FLIP, like before,
proving that the unbiasedness of the NIRL estimator is robust to our regularization scheme as well. One 
important result our validation work shows is that the NIRL-FLIP reconstruction for a wide range of features,
gives propagated error bars that are narrower when our FLIP regularization is applied.
This means that the template vs NIRL separation 
in $k$ regions of interest can be several 
$\sigma$ apart, which is a statistically significant improvement in reduction of lensing bias. Given that
this is a stringent validation test using below cosmic variance lensing contribution of the features (aka "weak" features),
using cosmic variance limited error bars, this provides strong motivation to use NIRL-FLIP for future model 
independent PPS reconstruction efforts and joint parameter estimation.

We also present the $\chi^2$ histories for both features, only in the NIRL-FLIP case,
and observe that the FLIP filter suppresses the late iteration $\chi^2$ rise, and in 
fact the unfiltered $\chi^2$ history actually crosses above the FLIP filtered $\chi^2$ 
at late iterations, which indicates the FLIP filter is suppressing noise overfitting to some extent,
which is an unexpected but welcome consequence of this regularization scheme.

We further look at the Fourier spectra of the final reconstructions for both features 
in figures~\ref{fig:fftboot_ft_wp} and \ref{fig:fftboot_ft_bump}. We can clearly see that 
flat spectrum like floor at high $\xi$ modes is suppressed and the NIRL estimator is 
closer to the true PPS feature spectrum than the template, by a comfortable order 
of magnitude in the case of the Gaussian bump feature. Not only are high $\xi$ "noisy" modes suppressed,
but the low $\xi$ mode excess (power above the feature spectrum) is also reduced for NIRL relative to template.
This suggests that the combination of NIRL debiaser and the FLIP filter reinforce each other to reduce 
both bias and noise leakage in our reconstructed PPS. 

% Keep all result figures before the concluding section.
\finishresultfloats
\section{Conclusions}
\label{sec:discuss}

In this work we present two novel updates to the Richardson Lucy deconvolution estimator
used in the context of reconstructing the inflationary scalar primordial power spectrum 
from lensed CMB temperature data.

The weak lensing effect on the CMB power spectrum adds a non-linear term to the deconvolution problem
that leads to a bias in naive reconstruction.
We propose an iterative lensing bias correcting update to the MRL estimator, which we call NIRL
and it is shown to be a robust debiasing estimator for a wide range of simulated features that 
probe a diverse PPS $k$ range and resolution, shown as a consistent reduction in the $k$ averaged,
absolute rms reconstruction error with respect to the debiased reconstruction of $0.21$--$1.18\%$ using NIRL,
compared with $0.97$--$3.24\%$ for fixed-template delensing, and a lower per $k$ rms error.
This lensing debiasing has the advantage of being truly model independent 
for free form PPS reconstruction, as well as avoiding spurious features arising from lensing power leakage 
in the PPS or other $\Lambda$CDM parameters in the case of future work on joint parameter estimation.

We validate NIRL further to a combined statistical significance evaluation by 
adding kernel sparsity using KSS from previous work, as well as cosmic variance limited error bars
and 1000 bootstrap realizations of the lensed $C_\ell^{TT}$ data for two representative features. 
We propose a regularization prescription, based on a first principle insight inspired by signal 
processing methods, specifically in our case, we are motivated by the delay transform method used in 21cm cosmology, 
to implement foreground avoidance and delay filtering, as well as the Fourier sensitivity of 
the physics of the radiative transfer kernel for CMB. We propose a Fourier low pass iterative filter
which we call FLIP, which is applied to the reconstructed PPS at each iteration after the first unfiltered $\chi^2$ minima
and converged on the second post FLIP filter $\chi^2$ minima. We show that it successfully suppresses 
noise overfitting to a significant extent while maintaining the unbiasedness of the NIRL estimator.
NIRL-FLIP gives $k$ averaged rms errors of $4.35\%$ for the
wavepacket and $2.07\%$ for the bump, versus $5.72\%$ and $6.11\%$ for
template-FLIP, with respect to the truth PPS. We get multi-$\sigma$ significance levels at key feature 
dominated $k$ regions.

The combined results provide a better than expected outcome. The Fourier spectra show that 
the NIRL-FLIP estimator jointly suppress noise overfitting at both low and high $\xi$ modes,
leading to greater precision and accuracy in the reconstructed PPS and its debiasing. It also is able 
to partially suppress noise overfitting with increasing iterations, with FLIP $\chi^2$ levels 
crossing below the unfiltered $\chi^2$ levels at late iterations. 

This motivates further application of these methods for future model independent PPS reconstruction efforts 
and joint parameter estimation. We should note that this work is one of the many methods that can be
used for PPS reconstruction amidst many other model based or model independent methods, which may have their 
own advantages or otherwise, in context of existing data and future data.

We expect to implement further improvements to the NIRL-FLIP estimator, including a more 
consistent exploration of the FLIP filter cutoff and stopping criteria. This work is nearing completion
and will be applied on the most recent CMB power spectrum data releases to obtain increasingly stringent bounds 
on model independent PPS reconstructions and therefore joint parameter estimation. We expect to report
promising results soon in a future publication.
We also expect to extend our methods to polarization and lensing reconstruction kernels and increase the statistical significance
of any PPS constraints.

\begin{acknowledgments}
  We would like to acknowledge Saurabh Singh and Shabbir Shaikh for useful discussions. 
This work has been carried out with the support of the Council of Scientific and Industrial Research (CSIR) 
and University Grants Commission (UGC) funding at Raman Research Institute (RRI), Bengaluru,
and Inter-University Centre for Astronomy and Astrophysics (IUCAA), Pune.
RSC acknowledges support from Raman Research Institute (RRI), Bengaluru.
DM acknowledges support from Raman Research Institute (RRI), Bengaluru.
TS acknowledges support from the J. C. Bose Grant of Anusandhan National Research Foundation, India.
The work acknowledges the use of the RRI High Performance Computing facility 
and the IUCAA High Performance Computing facility. 
\end{acknowledgments}


\begin{thebibliography}{99}

\bibitem{Starobinsky:1992ts}
A.~A.~Starobinsky,
``Spectrum of adiabatic perturbations in the universe when there are
singularities in the inflation potential,''
JETP Lett. \textbf{55}, 489--494 (1992).
\href{https://inspirehep.net/literature/343958}{INSPIRE record}.

\bibitem{JoySahniStarobinsky2008}
M.~Joy, V.~Sahni and A.~A.~Starobinsky,
``New universal local feature in the inflationary perturbation spectrum,''
Phys. Rev. D \textbf{77}, 023514 (2008),
\href{https://doi.org/10.1103/PhysRevD.77.023514}{doi:10.1103/PhysRevD.77.023514}
[arXiv:0711.1585 [astro-ph]].

\bibitem{ContaldiEtAl2003}
C.~R.~Contaldi, M.~Peloso, L.~Kofman and A.~Linde,
``Suppressing the lower Multipoles in the CMB Anisotropies,''
JCAP \textbf{07}, 002 (2003),
\href{https://doi.org/10.1088/1475-7516/2003/07/002}{doi:10.1088/1475-7516/2003/07/002}
[arXiv:astro-ph/0303636].

\bibitem{JainEtAl2009}
R.~K.~Jain, P.~Chingangbam, J.-O.~Gong, L.~Sriramkumar and T.~Souradeep,
``Punctuated inflation and the low CMB multipoles,''
JCAP \textbf{01}, 009 (2009),
\href{https://doi.org/10.1088/1475-7516/2009/01/009}{doi:10.1088/1475-7516/2009/01/009}
[arXiv:0809.3915 [astro-ph]].

\bibitem{Stewart1997RunningMass}
E.~D.~Stewart,
``Flattening the Inflaton's Potential with Quantum Corrections II,''
Phys. Rev. D \textbf{56}, 2019--2023 (1997),
\href{https://doi.org/10.1103/PhysRevD.56.2019}{doi:10.1103/PhysRevD.56.2019}
[arXiv:hep-ph/9703232].

\bibitem{CoviEtAl2004RunningMass}
L.~Covi, D.~H.~Lyth, A.~Melchiorri and C.~J.~Odman,
``The running-mass inflation model and WMAP,''
Phys. Rev. D \textbf{70}, 123521 (2004),
\href{https://doi.org/10.1103/PhysRevD.70.123521}{doi:10.1103/PhysRevD.70.123521}
[arXiv:astro-ph/0408129].

\bibitem{BarnabyEtAl2009}
N.~Barnaby, Z.~Huang, L.~Kofman and D.~Pogosyan,
``Cosmological Fluctuations from Infra-Red Cascading During Inflation,''
Phys. Rev. D \textbf{80}, 043501 (2009),
\href{https://doi.org/10.1103/PhysRevD.80.043501}{doi:10.1103/PhysRevD.80.043501}
[arXiv:0902.0615 [hep-th]].

\bibitem{RomanoSasaki2008}
A.~E.~Romano and M.~Sasaki,
``Effects of particle production during inflation,''
Phys. Rev. D \textbf{78}, 103522 (2008),
\href{https://doi.org/10.1103/PhysRevD.78.103522}{doi:10.1103/PhysRevD.78.103522}
[arXiv:0809.5142 [gr-qc]].

\bibitem{Hazra:2014goa}
D. K. Hazra, A. Shafieloo, G. F. Smoot and A. A. Starobinsky,
``Wiggly Whipped Inflation,''
JCAP 08, 048 (2014),
\href{https://doi.org/10.1088/1475-7516/2014/08/048}{doi:10.1088/1475-7516/2014/08/048}
[arXiv:1405.2012 [astro-ph.CO]].

\bibitem{AdamsCresswellEasther2001}
J.~A.~Adams, B.~Cresswell and R.~Easther,
``Inflationary perturbations from a potential with a step,''
Phys. Rev. D \textbf{64}, 123514 (2001),
\href{https://doi.org/10.1103/PhysRevD.64.123514}{doi:10.1103/PhysRevD.64.123514}
[arXiv:astro-ph/0102236].

\bibitem{ChenEastherLim2008}
X.~Chen, R.~Easther and E.~A.~Lim,
``Generation and Characterization of Large Non-Gaussianities in Single Field
Inflation,''
JCAP \textbf{04}, 010 (2008),
\href{https://doi.org/10.1088/1475-7516/2008/04/010}{doi:10.1088/1475-7516/2008/04/010}
[arXiv:0801.3295 [astro-ph]].

\bibitem{FlaugerEtAl2010}
R.~Flauger, L.~McAllister, E.~Pajer, A.~Westphal and G.~Xu,
``Oscillations in the CMB from Axion Monodromy Inflation,''
JCAP \textbf{06}, 009 (2010),
\href{https://doi.org/10.1088/1475-7516/2010/06/009}{doi:10.1088/1475-7516/2010/06/009}
[arXiv:0907.2916 [hep-th]].

\bibitem{AchucarroEtAl2011}
A.~Ach\'ucarro, J.-O.~Gong, S.~Hardeman, G.~A.~Palma and S.~P.~Patil,
``Features of heavy physics in the CMB power spectrum,''
JCAP \textbf{01}, 030 (2011),
\href{https://doi.org/10.1088/1475-7516/2011/01/030}{doi:10.1088/1475-7516/2011/01/030}
[arXiv:1010.3693 [hep-ph]].

\bibitem{MirandaHuAdshead2012}
V.~Miranda, W.~Hu and P.~Adshead,
``Warp Features in DBI Inflation,''
Phys. Rev. D \textbf{86}, 063529 (2012),
\href{https://doi.org/10.1103/PhysRevD.86.063529}{doi:10.1103/PhysRevD.86.063529}
[arXiv:1207.2186 [astro-ph.CO]].

\bibitem{BeanEtAl2008}
R.~Bean, X.~Chen, G.~Hailu, S.-H.~H.~Tye and J.~Xu,
``Duality Cascade in Brane Inflation,''
JCAP \textbf{03}, 026 (2008),
\href{https://doi.org/10.1088/1475-7516/2008/03/026}{doi:10.1088/1475-7516/2008/03/026}
[arXiv:0802.0491 [hep-th]].

\bibitem{Domenech:2019cyh}
G. Dom\`enech and M. Kamionkowski,
``Lensing anomaly and oscillations in the primordial power spectrum,''
JCAP 11, 040 (2019),
\href{https://doi.org/10.1088/1475-7516/2019/11/040}{doi:10.1088/1475-7516/2019/11/040}
[arXiv:1905.04323 [astro-ph.CO]].

\bibitem{Beutler:2019ojk}
F.~Beutler, M.~Biagetti, D.~Green, A.~Slosar and B.~Wallisch,
``Primordial Features from Linear to Nonlinear Scales,''
Phys. Rev. Res. \textbf{1}, 033209 (2019),
\href{https://doi.org/10.1103/PhysRevResearch.1.033209}{doi:10.1103/PhysRevResearch.1.033209}
[arXiv:1906.08758 [astro-ph.CO]].

\bibitem{Polarski:1992dq}
D.~Polarski and A.~A.~Starobinsky,
``Spectra of perturbations produced by double inflation with an
intermediate matter-dominated stage,''
Nucl. Phys. B \textbf{385}, 623--650 (1992),
\href{https://doi.org/10.1016/0550-3213(92)90062-G}{doi:10.1016/0550-3213(92)90062-G}.

\bibitem{AichEtAl2013}
M.~Aich, D.~K.~Hazra, L.~Sriramkumar and T.~Souradeep,
``Oscillations in the inflaton potential: Complete numerical treatment and
comparison with the recent and forthcoming CMB datasets,''
Phys. Rev. D \textbf{87}, 083526 (2013),
\href{https://doi.org/10.1103/PhysRevD.87.083526}{doi:10.1103/PhysRevD.87.083526}
[arXiv:1106.2798 [astro-ph.CO]].

\bibitem{GaoLangloisMizuno2012}
X.~Gao, D.~Langlois and S.~Mizuno,
``Influence of heavy modes on perturbations in multiple field inflation,''
JCAP \textbf{10}, 040 (2012),
\href{https://doi.org/10.1088/1475-7516/2012/10/040}{doi:10.1088/1475-7516/2012/10/040}
[arXiv:1205.5275 [hep-th]].

\bibitem{ChenNamjooWang2015}
X.~Chen, M.~H.~Namjoo and Y.~Wang,
``Models of the Primordial Standard Clock,''
JCAP \textbf{02}, 027 (2015),
\href{https://doi.org/10.1088/1475-7516/2015/02/027}{doi:10.1088/1475-7516/2015/02/027}
[arXiv:1411.2349 [astro-ph.CO]].

\bibitem{SlosarEtAl2019Features}
A.~Slosar, X.~Chen, C.~Dvorkin, D.~Green, P.~D.~Meerburg,
E.~Silverstein and B.~Wallisch,
``Scratches from the Past: Inflationary Archaeology through Features in the
Power Spectrum of Primordial Fluctuations,''
Astro2020 science white paper (2019)
[\href{https://arxiv.org/abs/1903.09883}{arXiv:1903.09883} [astro-ph.CO]].

\bibitem{CoviEtAl2006Step}
L.~Covi, J.~Hamann, A.~Melchiorri, A.~Slosar and I.~Sorbera,
``Inflation and WMAP three year data: Features have a Future!,''
Phys. Rev. D \textbf{74}, 083509 (2006),
\href{https://doi.org/10.1103/PhysRevD.74.083509}{doi:10.1103/PhysRevD.74.083509}
[arXiv:astro-ph/0606452].

\bibitem{HinshawEtAl2013WMAP9}
G. Hinshaw et al.,
``Nine-Year Wilkinson Microwave Anisotropy Probe (WMAP) Observations:
Cosmological Parameter Results,''
Astrophys. J. Suppl. 208, 19 (2013),
\href{https://doi.org/10.1088/0067-0049/208/2/19}{doi:10.1088/0067-0049/208/2/19}
[arXiv:1212.5226 [astro-ph.CO]].

\bibitem{Aghanim:2018eyx}
N. Aghanim et al. [Planck],
``Planck 2018 results. VI. Cosmological parameters,''
Astron. Astrophys. 641, A6 (2020),
\href{https://doi.org/10.1051/0004-6361/201833910}{doi:10.1051/0004-6361/201833910}
[arXiv:1807.06209 [astro-ph.CO]].

\bibitem{Akrami:2018odb}
Y. Akrami et al. [Planck],
``Planck 2018 results. X. Constraints on inflation,''
Astron. Astrophys. 641, A10 (2020),
\href{https://doi.org/10.1051/0004-6361/201833887}{doi:10.1051/0004-6361/201833887}
[arXiv:1807.06211 [astro-ph.CO]].

\bibitem{Rosenberg:2022sdy}
E. Rosenberg, S. Gratton and G. Efstathiou,
``CMB power spectra and cosmological parameters from Planck PR4 with
CamSpec,''
Mon. Not. Roy. Astron. Soc. 517, 4620--4636 (2022),
\href{https://doi.org/10.1093/mnras/stac2744}{doi:10.1093/mnras/stac2744}
[arXiv:2205.10869 [astro-ph.CO]].

\bibitem{LouisEtAl2025ACTDR6}
T. Louis et al. [Atacama Cosmology Telescope],
``The Atacama Cosmology Telescope: DR6 power spectra, likelihoods and
$\Lambda$CDM parameters,''
JCAP 11, 062 (2025),
\href{https://doi.org/10.1088/1475-7516/2025/11/062}{doi:10.1088/1475-7516/2025/11/062}
[arXiv:2503.14452 [astro-ph.CO]].

\bibitem{CamphuisEtAl2026SPT3GD1}
E. Camphuis et al. [SPT-3G],
``SPT-3G D1: CMB temperature and polarization power spectra and cosmology from
2019 and 2020 observations of the SPT-3G main field,''
Phys. Rev. D 113, 083504 (2026),
\href{https://doi.org/10.1103/7wt3-9v2y}{doi:10.1103/7wt3-9v2y}
[arXiv:2506.20707 [astro-ph.CO]].

\bibitem{HazraEtAl2016PlanckPolarization}
D. K. Hazra, A. Shafieloo, G. F. Smoot and A. A. Starobinsky,
``Primordial features and Planck polarization,''
JCAP 09, 009 (2016),
\href{https://doi.org/10.1088/1475-7516/2016/09/009}{doi:10.1088/1475-7516/2016/09/009}
[arXiv:1605.02106 [astro-ph.CO]].

\bibitem{HazraEtAl2021InflationStory}
D. K. Hazra, D. Paoletti, I. Debono, A. Shafieloo, G. F. Smoot and A. A. Starobinsky,
``Inflation Story: slow-roll and beyond,''
JCAP 12, 038 (2021),
\href{https://doi.org/10.1088/1475-7516/2021/12/038}{doi:10.1088/1475-7516/2021/12/038}
[arXiv:2107.09460 [astro-ph.CO]].

\bibitem{MeerburgEtAl2014PlanckOscillations}
P. D. Meerburg, D. N. Spergel and B. D. Wandelt,
``Searching for oscillations in the primordial power spectrum. II. Constraints
from Planck data,''
Phys. Rev. D 89, 063537 (2014),
\href{https://doi.org/10.1103/PhysRevD.89.063537}{doi:10.1103/PhysRevD.89.063537}
[arXiv:1308.3705 [astro-ph.CO]].

\bibitem{AntonyEtAl2026PlanckSPTOscillations}
A. Antony, F. Finelli, D. K. Hazra, D. Paoletti and A. Shafieloo,
``A search for super-imposed oscillations to the primordial power spectrum in
Planck and SPT-3G 2018 data,''
JCAP 03, 029 (2026),
\href{https://doi.org/10.1088/1475-7516/2026/03/029}{doi:10.1088/1475-7516/2026/03/029}
[arXiv:2403.19575 [astro-ph.CO]].

\bibitem{RaffaelliBallardini2025Knots}
A. Raffaelli and M. Ballardini,
``Knot reconstruction of the scalar primordial power spectrum with Planck,
ACT, and SPT CMB data,''
JCAP 07, 077 (2025),
\href{https://doi.org/10.1088/1475-7516/2025/07/077}{doi:10.1088/1475-7516/2025/07/077}
[arXiv:2503.10609].

\bibitem{BridgesEtAl2009PPS}
M. Bridges, F. Feroz, M. P. Hobson and A. N. Lasenby,
``Bayesian optimal reconstruction of the primordial power spectrum,''
Mon. Not. Roy. Astron. Soc. 400, 1075--1084 (2009),
\href{https://doi.org/10.1111/j.1365-2966.2009.15525.x}{doi:10.1111/j.1365-2966.2009.15525.x}
[arXiv:0812.3541].

\bibitem{AslanyanEtAl2014Knots}
G. Aslanyan, L. C. Price, K. N. Abazajian and R. Easther,
``The Knotted Sky I: Planck constraints on the primordial power spectrum,''
JCAP 08, 052 (2014),
\href{https://doi.org/10.1088/1475-7516/2014/08/052}{doi:10.1088/1475-7516/2014/08/052}
[arXiv:1403.5849].

\bibitem{HandleyEtAl2019Reconstruction}
W. J. Handley, A. N. Lasenby, H. V. Peiris and M. P. Hobson,
``Bayesian inflationary reconstructions from Planck 2018 data,''
Phys. Rev. D 100, 103511 (2019),
\href{https://doi.org/10.1103/PhysRevD.100.103511}{doi:10.1103/PhysRevD.100.103511}
[arXiv:1908.00906].

\bibitem{SealfonEtAl2005Splines}
C. Sealfon, L. Verde and R. Jimenez,
``Smoothing spline primordial power spectrum reconstruction,''
Phys. Rev. D 72, 103520 (2005),
\href{https://doi.org/10.1103/PhysRevD.72.103520}{doi:10.1103/PhysRevD.72.103520}
[arXiv:astro-ph/0506707].

\bibitem{VerdePeiris2008Splines}
L. Verde and H. V. Peiris,
``On minimally parametric primordial power spectrum reconstruction and the
evidence for a red tilt,''
JCAP 07, 009 (2008),
\href{https://doi.org/10.1088/1475-7516/2008/07/009}{doi:10.1088/1475-7516/2008/07/009}
[arXiv:0802.1219].

\bibitem{TocchiniEtAl2005PPS}
D. Tocchini-Valentini, M. Douspis and J. Silk,
``A new search for features in the primordial power spectrum,''
Mon. Not. Roy. Astron. Soc. 359, 31--35 (2005),
\href{https://doi.org/10.1111/j.1365-2966.2005.08771.x}{doi:10.1111/j.1365-2966.2005.08771.x}
[arXiv:astro-ph/0402583].

\bibitem{HuntSarkar2014Tikhonov}
P. Hunt and S. Sarkar,
``Reconstruction of the primordial power spectrum of curvature perturbations
using multiple data sets,''
JCAP 01, 025 (2014),
\href{https://doi.org/10.1088/1475-7516/2014/01/025}{doi:10.1088/1475-7516/2014/01/025}
[arXiv:1308.2317].

\bibitem{HuntSarkar2015Planck}
P. Hunt and S. Sarkar,
``Search for features in the spectrum of primordial perturbations using Planck
and other datasets,''
JCAP 12, 052 (2015),
\href{https://doi.org/10.1088/1475-7516/2015/12/052}{doi:10.1088/1475-7516/2015/12/052}
[arXiv:1510.03338].

\bibitem{GauthierBucher2012PPS}
C. Gauthier and M. Bucher,
``Reconstructing the primordial power spectrum from the CMB,''
JCAP 10, 050 (2012),
\href{https://doi.org/10.1088/1475-7516/2012/10/050}{doi:10.1088/1475-7516/2012/10/050}
[arXiv:1209.2147].

\bibitem{NagataYokoyama2009Bandpowers}
R. Nagata and J. Yokoyama,
``Band-power reconstruction of the primordial fluctuation spectrum by the
maximum likelihood reconstruction method,''
Phys. Rev. D 79, 043010 (2009),
\href{https://doi.org/10.1103/PhysRevD.79.043010}{doi:10.1103/PhysRevD.79.043010}
[arXiv:0812.4585].

\bibitem{MatsumiyaEtAl2002Inversion}
M. Matsumiya, M. Sasaki and J. Yokoyama,
``Cosmic inversion: Reconstructing the primordial spectrum from CMB
anisotropy,''
Phys. Rev. D 65, 083007 (2002),
\href{https://doi.org/10.1103/PhysRevD.65.083007}{doi:10.1103/PhysRevD.65.083007}
[arXiv:astro-ph/0111549].

\bibitem{MatsumiyaEtAl2003Inversion}
M. Matsumiya, M. Sasaki and J. Yokoyama,
``Cosmic inversion. II. An iterative method for reproducing the primordial
spectrum from the CMB data,''
JCAP 02, 003 (2003),
\href{https://doi.org/10.1088/1475-7516/2003/02/003}{doi:10.1088/1475-7516/2003/02/003}
[arXiv:astro-ph/0210365].

\bibitem{KogoEtAl2004Inversion}
N. Kogo, M. Matsumiya, M. Sasaki and J. Yokoyama,
``Reconstructing the Primordial Spectrum from WMAP Data by the Cosmic
Inversion Method,''
Astrophys. J. 607, 32--39 (2004),
\href{https://doi.org/10.1086/383339}{doi:10.1086/383339}
[arXiv:astro-ph/0309662].

\bibitem{2010JCAP...01..016N}
G. Nicholson, C. R. Contaldi and P. Paykari,
``Reconstruction of the primordial power spectrum by direct inversion,''
JCAP 01, 016 (2010),
\href{https://doi.org/10.1088/1475-7516/2010/01/016}{doi:10.1088/1475-7516/2010/01/016}
[arXiv:0909.5092].

\bibitem{Leach2006PCA}
S. M. Leach,
``Measuring the primordial power spectrum: principal component analysis of the
cosmic microwave background,''
Mon. Not. Roy. Astron. Soc. 372, 646--654 (2006),
\href{https://doi.org/10.1111/j.1365-2966.2006.10842.x}{doi:10.1111/j.1365-2966.2006.10842.x}
[arXiv:astro-ph/0506390].

\bibitem{MukherjeeWang2003Wavelets}
P. Mukherjee and Y. Wang,
``Direct Wavelet Expansion of the Primordial Power Spectrum,''
Astrophys. J. 598, 779--784 (2003),
\href{https://doi.org/10.1086/379107}{doi:10.1086/379107}
[arXiv:astro-ph/0301562].

\bibitem{2014A&A...566A..77P}
P. Paykari, F. Lanusse, J.-L. Starck, F. Sureau and J. Bobin,
``PRISM: Sparse recovery of the primordial power spectrum,''
Astron. Astrophys. 566, A77 (2014),
\href{https://doi.org/10.1051/0004-6361/201322326}{doi:10.1051/0004-6361/201322326}
[arXiv:1402.1983].

\bibitem{Lanusse:2014sra}
F. Lanusse, P. Paykari, J.-L. Starck, F. Sureau, J. Bobin and A. Rassat,
``PRISM: Recovery of the primordial spectrum from Planck data,''
Astron. Astrophys. 571, L1 (2014),
\href{https://doi.org/10.1051/0004-6361/201424420}{doi:10.1051/0004-6361/201424420}
[arXiv:1410.2571].

\bibitem{ChakiEtAl2025GP}
S. Chaki, A. Nicola, A. Spurio Mancini, D. Piras and R. Reischke,
``Constraining the primordial power spectrum using a differentiable
likelihood,''
JCAP 08, 068 (2025),
\href{https://doi.org/10.1088/1475-7516/2025/08/068}{doi:10.1088/1475-7516/2025/08/068}
[arXiv:2503.00108].

\bibitem{LodhaEtAl2024GA}
K. Lodha, L. Pinol, S. Nesseris, A. Shafieloo, W. Sohn and M. Fasiello,
``Searching for local features in primordial power spectrum using genetic
algorithms,''
Mon. Not. Roy. Astron. Soc. 530, 1424--1435 (2024),
\href{https://doi.org/10.1093/mnras/stae803}{doi:10.1093/mnras/stae803}
[arXiv:2308.04940].

\bibitem{1972JOSA...62...55R} Richardson W.~H., 1972, JOSA, 62, 55
\href{https://doi.org/10.1364/JOSA.62.000055}{doi:10.1364/JOSA.62.000055}.

\bibitem{1974AJ.....79..745L} Lucy L.~B., 1974, AJ, 79, 745. \href{https://doi.org/10.1086/111605}{doi:10.1086/111605}

\bibitem{1992AJ....104.1260L}
L.~B.~Lucy,
``Resolution limits for deconvolved images,''
Astron. J. \textbf{104}, 1260--1265 (1992),
\href{https://doi.org/10.1086/116315}{doi:10.1086/116315}.


\bibitem{SheppVardi1982MLE}
Shepp, L. \& Vardi, Y. Maximum Likelihood Reconstruction for Emission Tomography. 
{\em IEEE Transactions On Medical Imaging}. 
\textbf{1}, 113-122 (1982)
\href{https://ieeexplore.ieee.org/stamp/stamp.jsp?tp=&arnumber=4307558}{doi:10.1109/TMI.1982.4307558}


\bibitem{1994A&A...289..983L}
L.~B.~Lucy,
``Optimum strategies for inverse problems in statistical astronomy,''
Astron. Astrophys. \textbf{289}, 983--994 (1994).
\href{https://ui.adsabs.harvard.edu/abs/1994A\%26A...289..983L/abstract}{NASA ADS record}.

\bibitem{DeyEtAl2006RLTV}
N.~Dey, L.~Blanc-F\'eraud, C.~Zimmer, P.~Roux, Z.~Kam, J.-C.~Olivo-Marin and J.~Zerubia,
``Richardson--Lucy algorithm with total variation regularization for 3D confocal microscope deconvolution,''
Microsc. Res. Tech. \textbf{69}, 260--266 (2006),
\href{https://doi.org/10.1002/jemt.20294}{doi:10.1002/jemt.20294}.

\bibitem{SchmittEtAl2012PoissonSphere}
J.~Schmitt, J.-L.~Starck, J.-M.~Casandjian, J.~Fadili and I.~Grenier,
``Multichannel Poisson denoising and deconvolution on the sphere: Application to the Fermi Gamma-ray Space Telescope,''
Astron. Astrophys. \textbf{546}, A114 (2012),
\href{https://doi.org/10.1051/0004-6361/201118234}{doi:10.1051/0004-6361/201118234}
[arXiv:1206.2787 [astro-ph.IM]].

\bibitem{1993MNRAS.265..145B} Baugh, C.~M., Efstathiou, G.\ 1993.\ The three-dimensional power spectrum measured from the APM galaxy
survey - I. Use of the angular correlation function..\ Monthly Notices of the Royal Astronomical
Society 265, 145--156.
\href{https://doi.org/10.1093/mnras/265.1.145}{doi:10.1093/mnras/265.1.145}.

\bibitem{1994MNRAS.267..323B} Baugh, C.~M., Efstathiou, G.\ 1994.\ The three-dimensional power spectrum measured from the APM Galaxy
Survey-2. Use of the two-dimensional power spectrum.\ Monthly Notices of the Royal Astronomical Society 267,
323--332.
\href{https://doi.org/10.1093/mnras/267.2.323}{doi:10.1093/mnras/267.2.323}.

\bibitem{Shafieloo:2003gf}
A. Shafieloo and T. Souradeep,
``Primordial power spectrum from WMAP,''
Phys. Rev. D 70, 043523 (2004),
\href{https://doi.org/10.1103/PhysRevD.70.043523}{doi:10.1103/PhysRevD.70.043523}
[arXiv:astro-ph/0312174].

\bibitem{Shafieloo:2006hs}
A.~Shafieloo, T.~Souradeep, P.~Manimaran, P.~K.~Panigrahi and R.~Rangarajan,
``Features in the Primordial Spectrum from WMAP: A Wavelet Analysis,''
Phys. Rev. D \textbf{75} (2007), 123502
\href{https://doi.org/10.1103/PhysRevD.75.123502}{doi:10.1103/PhysRevD.75.123502}
[arXiv:astro-ph/0611352 [astro-ph]].

\bibitem{Nicholson:2009pi}
G.~Nicholson and C.~R.~Contaldi,
``Reconstruction of the Primordial Power Spectrum using Temperature and Polarisation Data from Multiple Experiments,''
JCAP \textbf{07}, 011 (2009)
\href{https://doi.org/10.1088/1475-7516/2009/07/011}{doi:10.1088/1475-7516/2009/07/011}
[arXiv:0903.1106 [astro-ph.CO]].

\bibitem{SohnEtAl2022PPS}
W. Sohn, A. Shafieloo and D. K. Hazra,
``Deblurring the early Universe: reconstruction of primordial power spectrum
from Planck CMB using image analysis techniques,''
JCAP 03, 056 (2024),
\href{https://doi.org/10.1088/1475-7516/2024/03/056}{doi:10.1088/1475-7516/2024/03/056}
[arXiv:2211.15139 [astro-ph.CO]].

\bibitem{Shafieloo:2007tk}
A.~Shafieloo and T.~Souradeep,
``Estimation of Primordial Spectrum with post-WMAP 3 year data,''
Phys. Rev. D \textbf{78}, 023511 (2008)
\href{https://doi.org/10.1103/PhysRevD.78.023511}{doi:10.1103/PhysRevD.78.023511}
[arXiv:0709.1944 [astro-ph]].

\bibitem{2010JCAP...04..010H}
J. Hamann, A. Shafieloo and T. Souradeep,
``Features in the primordial power spectrum? A frequentist analysis,''
JCAP 04, 010 (2010),
\href{https://doi.org/10.1088/1475-7516/2010/04/010}{doi:10.1088/1475-7516/2010/04/010}
[arXiv:0912.2728 [astro-ph.CO]].

\bibitem{2011NJPh...13j3024S} Shafieloo, A., Souradeep, T.\ 2011.\ Assumptions of the primordial spectrum and cosmological parameter estimation.\
New Journal of Physics 13.
\href{https://doi.org/10.1088/1367-2630/13/10/103024}{doi:10.1088/1367-2630/13/10/103024}.

\bibitem{Hazra:2013xva}
D.~K.~Hazra, A.~Shafieloo and T.~Souradeep,
``Primordial power spectrum: a complete analysis with the WMAP nine-year data,''
JCAP \textbf{07}, 031 (2013)
\href{https://doi.org/10.1088/1475-7516/2013/07/031}{doi:10.1088/1475-7516/2013/07/031}
[arXiv:1303.4143 [astro-ph.CO]].

\bibitem{Hazra:2013eva}
D.~K.~Hazra, A.~Shafieloo and T.~Souradeep,
``Cosmological parameter estimation with free-form primordial power spectrum,''
Phys. Rev. D \textbf{87} (2013) no.12, 123528
\href{https://doi.org/10.1103/PhysRevD.87.123528}{doi:10.1103/PhysRevD.87.123528}
[arXiv:1303.5336 [astro-ph.CO]].

\bibitem{HazraEtAl2024PlanckACT}
D.~K.~Hazra, B.~Beringue, J.~Errard, A.~Shafieloo and G.~F.~Smoot,
``Exploring the discrepancy between Planck PR3 and ACT DR4,''
JCAP \textbf{12}, 038 (2024),
\href{https://doi.org/10.1088/1475-7516/2024/12/038}{doi:10.1088/1475-7516/2024/12/038}
[arXiv:2406.06296 [astro-ph.CO]].

\bibitem{ChandraEtAl2026PPS}
D. Chandra, D. K. Hazra, A. Shafieloo and T. Souradeep,
``The Primordial power spectrum from the largest to smallest CMB scales,''
Preprint, version 1 (2026),
[\href{https://arxiv.org/abs/2608.24413}{arXiv:2608.24413} [astro-ph.CO]].

\bibitem{Hazra:2014jwa}
D. K. Hazra, A. Shafieloo and T. Souradeep,
``Primordial power spectrum from Planck,''
JCAP 11, 011 (2014),
\href{https://doi.org/10.1088/1475-7516/2014/11/011}{doi:10.1088/1475-7516/2014/11/011}
[arXiv:1406.4827 [astro-ph.CO]].

\bibitem{Hazra:2018opk}
D. K. Hazra, A. Shafieloo and T. Souradeep,
``Parameter discordance in Planck CMB and low-redshift measurements:
projection in the primordial power spectrum,''
JCAP 04, 036 (2019),
\href{https://doi.org/10.1088/1475-7516/2019/04/036}{doi:10.1088/1475-7516/2019/04/036}
[arXiv:1810.08101 [astro-ph.CO]].

\bibitem{Keeley:2020rmo}
R. E. Keeley, A. Shafieloo, D. K. Hazra and T. Souradeep,
``Inflation Wars: A New Hope,''
JCAP 09, 055 (2020),
\href{https://doi.org/10.1088/1475-7516/2020/09/055}{doi:10.1088/1475-7516/2020/09/055}
[arXiv:2006.12710 [astro-ph.CO]].

\bibitem{HazraAntonyShafieloo2022OneSpectrum}
D. K. Hazra, A. Antony and A. Shafieloo,
``One spectrum to cure them all: Signature from early Universe solves major
anomalies and tensions in cosmology,''
JCAP 08, 063 (2022),
\href{https://doi.org/10.1088/1475-7516/2022/08/063}{doi:10.1088/1475-7516/2022/08/063}
[arXiv:2201.12000 [astro-ph.CO]].

\bibitem{Chandra:2021ydm}
R.~S.~Chandra and T.~Souradeep,
JCAP \textbf{10}, 081 (2021)
\href{https://doi.org/10.1088/1475-7516/2021/10/081}{doi:10.1088/1475-7516/2021/10/081}
[arXiv:2104.12253 [astro-ph.CO]].

\bibitem{RLnature2025}
Y. Liu, S. Panezai, Y. Wang and S. Stallinga,
``Noise amplification and ill-convergence of Richardson-Lucy deconvolution,''
Nat. Commun. 16, 911 (2025).
\href{https://doi.org/10.1038/s41467-025-56241-x}{doi:10.1038/s41467-025-56241-x}.
% [].

% \bibitem{Aghanim:2018oex}
% N.~Aghanim \textit{et al.} [Planck],
% ``Planck 2018 results. VIII. Gravitational lensing,''
% Astron. Astrophys. \textbf{641} (2020), A8
% \href{https://doi.org/10.1051/0004-6361/201833886}{doi:10.1051/0004-6361/201833886}
% [arXiv:1807.06210 [astro-ph.CO]].

\bibitem{Hu:2000ee}
W.~Hu,
``Weak lensing of the CMB: A harmonic approach,''
Phys. Rev. D \textbf{62} (2000), 043007
\href{https://doi.org/10.1103/PhysRevD.62.043007}{doi:10.1103/PhysRevD.62.043007}
[arXiv:astro-ph/0001303 [astro-ph]].

\bibitem{Lewis:2006fu}
A.~Lewis and A.~Challinor,
``Weak gravitational lensing of the CMB,''
Phys. Rept. \textbf{429} (2006), 1-65
\href{https://doi.org/10.1016/j.physrep.2006.03.002}{doi:10.1016/j.physrep.2006.03.002}
[arXiv:astro-ph/0601594 [astro-ph]].

\bibitem{2011PhRvD..83d3005H} Hanson, D., Challinor, A., Efstathiou, G., Bielewicz, P.\ 2011.\ CMB temperature lensing power reconstruction.\
Physical Review D 83.
\href{https://doi.org/10.1103/PhysRevD.83.043005}{doi:10.1103/PhysRevD.83.043005}.

\bibitem{Lewis:1999bs}
A.~Lewis, A.~Challinor and A.~Lasenby,
``Efficient computation of CMB anisotropies in closed FRW models,''
Astrophys. J. \textbf{538}, 473-476 (2000)
\href{https://doi.org/10.1086/309179}{doi:10.1086/309179}
[arXiv:astro-ph/9911177 [astro-ph]].
\url{https://camb.info/}

\bibitem{Aghanim:2019ame}
N.~Aghanim \textit{et al.} [Planck],
``Planck 2018 results. V. CMB power spectra and likelihoods,''
Astron. Astrophys. \textbf{641} (2020), A5
\href{https://doi.org/10.1051/0004-6361/201936386}{doi:10.1051/0004-6361/201936386}
[arXiv:1907.12875 [astro-ph.CO]].

\bibitem{Shaikh:2016hrf}
S.~Shaikh, S.~Mukherjee, A.~Rotti and T.~Souradeep,
``Constraining stochastic gravitational wave background from weak lensing of CMB B-modes,''
JCAP \textbf{09}, 029 (2016)
\href{https://doi.org/10.1088/1475-7516/2016/09/029}{doi:10.1088/1475-7516/2016/09/029}
[arXiv:1606.08862 [astro-ph.CO]].


\end{thebibliography}
\end{document}